\documentclass[DIV=calc,paper=a4,fontsize=11pt,twocolumn]{scrartcl} 

\usepackage[english]{babel}
\usepackage[protrusion=true,expansion=true]{microtype}
\usepackage{amsmath,amsfonts,amsthm}
\usepackage[final]{graphicx}
\usepackage{xcolor}
\usepackage[normal,small,hypcap,up,labelfont=bf,textfont=it]{caption}
\usepackage{epstopdf}
\usepackage{subfig}
\usepackage{booktabs}
\usepackage{fix-cm}
\usepackage{amssymb,amsfonts}
\usepackage{dsfont}
\usepackage{bbm}
\usepackage{pstricks}
\usepackage{cite}
\usepackage[utf8]{inputenc}
\usepackage[perpage,symbol*]{footmisc}
\usepackage[varg]{txfonts}
\usepackage{balance}
\usepackage{fancyhdr}
\PassOptionsToPackage{hyphens}{url}\usepackage[pdfencoding=auto,psdextra]{hyperref}
\usepackage{bookmark}
\usepackage{verbatim}
\usepackage{fontenc}
\usepackage{cuted}
\usepackage{braket}

\usepackage{bm}
\usepackage{mathrsfs}

\theoremstyle{definition}

\theoremstyle{plain}

\DeclareCaptionFont{mycolor}{\color[HTML]{000000}}
\usepackage{sectsty}													
\allsectionsfont{
\color[HTML]{31ADF3}\usefont{OT1}{phv}{b}{n}
}

\sectionfont{
\color[HTML]{31ADF3}\usefont{OT1}{phv}{b}{n}
}

\usepackage{fancyhdr}												
\usepackage{lettrine}
\newcommand{\initial}[1]{%
\lettrine[lines=3,lhang=0.3,nindent=0em]{
\color[HTML]{31ADF3}
{\textsf{#1}}}{}}

\usepackage{titling}															

\newcommand{\HorRule}{\color[HTML]{31ADF3}
\rule{\linewidth}{1pt}%
}

\pretitle{\vspace{-30pt} \begin{flushleft} \HorRule
\fontsize{34}{34} \usefont{OT1}{phv}{b}{n} \color[HTML]{31ADF3} \selectfont
}
\title{Quantum Cryptography:\\ Key Distribution and Beyond}					
\posttitle{\par\end{flushleft}\vskip 0.5em}

\preauthor{\begin{flushleft}\large \lineskip 0.5em \usefont{OT1}{phv}{b}{sl} \color[HTML]{31ADF3}}
\author{Akshata Shenoy-Hejamadi$^{~\mathsf{1}}$, Anirban Pathak$^{~\mathsf{2}}$ \& Srikanth Radhakrishna$^{~\mathsf{3}}$\\[8pt]}											
\postauthor{\footnotesize \usefont{OT1}{phv}{m}{sl} \color[HTML]{000000}
$^{\mathsf{1}}$ Group of Applied Physics, University of Geneva, Switzerland. E-mail: \href{mailto:akshata.shenoy@etu.unige.ch}{akshata.shenoy@etu.unige.ch}\\
$^{\mathsf{2}}$ Jaypee Institute of Information Technology, Noida, India. E-mail: \href{mailto:anirban.pathak@jiit.ac.in}{anirban.pathak@jiit.ac.in}\\
$^{\mathsf{3}}$ Poornaprajna Institute of Scientific Research, Bangalore, India. E-mail: \href{mailto:srik@poornaprajna.org}{srik@poornaprajna.org}\\[10pt]		
\scriptsize\usefont{OT1}{phv}{m}{n} \color[HTML]{31ADF3}{\textbf{Editors: \emph{Subhash Kak}, \emph{Tabish Qureshi} \& \emph{Danko Georgiev}} }\\[5pt]
\color[HTML]{000000}{Article history: Submitted on December 21, 2016;  Accepted on May 10, 2017; Published on June 18, 2017; Corrected on February 15, 2018.}
\par\end{flushleft}\HorRule}

\date{}																				

\begin{document}
\maketitle
\thispagestyle{fancy} 			
\initial{U}\textbf{niquely among the sciences, quantum cryptography has driven both
foundational research as well as practical real-life applications. We
review the progress of quantum cryptography in the last decade,
covering quantum key distribution and other applications.\\ Quanta 2017; 6: 1--47.}

\begin{figure}[b!]
\rule{245 pt}{0.5 pt}\\[3pt]
\raisebox{-0.2\height}{\includegraphics[width=5mm]{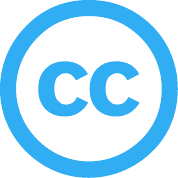}}\raisebox{-0.2\height}{\includegraphics[width=5mm]{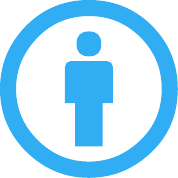}}
\footnotesize{This is an open access article distributed under the terms of the Creative Commons Attribution License \href{http://creativecommons.org/licenses/by/3.0/}{CC-BY-3.0}, which permits unrestricted use, distribution, and reproduction in any medium, provided the original author and source are credited.}
\end{figure}

\section{Introduction}

Cryptography is the technique of concealing confidential information
using physical or mathematical means. While cryptologists find newer
methods to conceal a secret, cryptanalysts devise powerful methods to
compromise the same. This recursive cat-and-mouse game has pushed the
field and driven progress in it tremendously, motivating the
participation of a large group of physicists, mathematicians and
engineers.

The seminal work of Peter W. Shor \cite{Sho94} uncovered the security threat
that quantum computation posed on all classical cryptographic schemes
that are based on computational assumptions, such as the hardness of
the discrete logarithm problem and the factorization problem. One such
cryptographic scheme is the Rivest--Shamir--Adleman (RSA) scheme, which
is widely used in \mbox{e-commerce} today. RSA in today's world is safe so
long as a scalable quantum computer remains unfeasible. However,
ideally, we wish to guarantee cryptographic security that follows only
from basic physics. This is the promise of quantum cryptography.

In particular, note the recent report on a loophole-free test of
Bell's inequality \cite{HBD+15}, thereby conclusively verifying the
existence of quantum nonlocality in Nature, and also attesting to the
advancement of experimental techniques to prepare, transmit,
manipulate and measure quantum information. Another
direction in cryptography is to provide practicable tools embracing
experimental limitations, e.g. quantum key distribution with pulses having mean photon
number much larger than one \cite{CRV+15}.

Several quantum cryptographic tools have now been
commercialized. \href{http://www.idquantique.com/}{ID-Quantique}, a major player in the quantum
cryptography industry, sells complete cryptographic solutions. Their
products include network encryption systems, quantum cryptographic
systems especially designed for industry and government, a
quantum random number generator, a state-of-art photon counting device,
single photon source, etc. QUANTIS, a quantum random number generator from
ID-Quantique deserves special mention, as it is used in quantum key distribution and
various quantum-classical hybrid machines (e.g., in casinos); the
CLAVIS series of products, which provide a platform for cryptography
research, are worth noting. Further, ID-Quantique's cryptographic
solution provides an open platform where buyers can incorporate their
own encryption algorithms.

Further, there are several other companies trying to commercialize quantum key distribution (see a long list of such companies at \href{https://en.wikipedia.org/wiki/List_of_Companies_involved_in_Quantum_Computing_or_Communication}{Wikipedia}, which depicts the importance of the field). Among this set of large number of other companies and the interesting
products developed by them, we would like to point out a few. \href{http://www.toshiba.eu/eu/Cambridge-Research-Laboratory/Quantum-Information-Group/Quantum-Key-Distribution/Toshiba-QKD-system/}{Toshiba}
markets an excellent room temperature single-photon
detector, a photon number resolving detector and a quantum key distribution system using
the T12 protocol \cite{T12}, in which the probability that bit values
are encoded in $X$ and $Z$ basis are different (otherwise, T12 is
similar to Bennett--Brassard 1984 protocol (BB84)) and decoy qubits are
used. A very attractive example of quantum-classical hybrid
cryptographic product is the world's first quantum-key-distribution-based one-time-pad
mobile phone software designed by \href{http://www.mitsubishielectric.com/company/rd/index.html}{Mitsubishi Electric}.

The interaction between academia and industry, and the development of
commercially viable products as a result, has been relatively thriving
in this area. In 2015, H.~Zbinden and his colleagues at GAP-Optique,
University of Geneva, performed a record breaking long distance quantum key distribution
experiment using a coherent one-way scheme that uses decoy qubits and
a variant of BB84. They successfully distributed the key in a secure
manner over 307 km. It took only a few months for the development of
the corresponding commercial product, as in October 2015, \href{http://www.idquantique.com/}{ID-Quantique}
introduced a commercial product using the same protocol (cf. Cerberis
QKD Blade at ID-Quantique).

While quantum key distribution remains the most popular
application of quantum cryptography, potential usefulness has been
recognized for other areas, in particular for \textit{distrustful
 cryptography}. This involves players with conflicting interests who do not necessarily trust one another, unlike in quantum key distribution. The present
review will try to cover many such areas, including relativistic
quantum cryptography, developed in the last decade since two
comprehensive reviews on quantum key distribution done in the previous decade
\cite{GRT+02,ABB+07}.

The present review is arranged as follows. In Section~\ref{sec:QKD},
we revisit quantum key distribution, briefly explaining intuitive and rigorous proofs of
security, presenting some variants of quantum key distribution going beyond BB84, among
them semi-quantum protocols, and touching on the issue of
composability, which is relevant for a large-scale implementation of a
quantum cryptography. A modification of the quantum key distribution, allowing for secure
deterministic communication, and other allied protocols, is discussed
in Section \ref{sec:qsdc}. In Section \ref{sec:CQC}, we cover the
paradigm of counterfactual key distribution, which is based on
interaction-free measurements. Subsequently, in Section
\ref{sec:DIS}, we discuss the practically important area of device
independence, in particular devoting subsections to the issues of side
channels, and then five classifications of device independence, namely
full, one-sided-, semi-, measurement- and detector-device
independence. The formalism of device independence can in principle
also be useful in a world where quantum mechanics fails to be valid, being replaced
by a non-signaling theory. We also briefly touch upon this, along
with the issue of self-testing, in the final subsection. Next, we
cover various other issues in cryptography besides quantum key distribution, covering
quantum versions for cryptotasks such as random number generation,
strong and weak coin tossing, private querying, secret sharing and
privacy preserving tasks. Some crypto-tasks not possible in
non-relativistic classical cryptography become feasible with the
inclusion of relativity or the conjunction of relativity and quantum mechanics. These
issues are discussed in Section \ref{sec:RC}. Technological issues
encountered in practical realization of quantum cryptography are
discussed in Section \ref{sec:tech}. After covering continuous
variable quantum cryptography in Section \ref{sec:cvqkd}, we conclude
in Section \ref{sec:conclu}.

\section{Quantum key distribution\label{sec:QKD}}

Quantum cryptography was born when S. Wiesner came up with the idea
of quantum money in the 1970s, though his paper eventually appeared
only in 1983. In 1984, Bennett and Brassard introduced their famous,
eponymous four-state protocol BB84 \cite{BB84}, using encoding based
on photon polarization. This was seminal in showing how quantum
features like uncertainty, impossibility of perfectly discriminating
non-orthogonal states and measurement disturbance were `just what the
doctor ordered' as far as secret communication goes. For the first
time, it became clear how quantum physical laws can provide
unconditional security, impossible classically. Since then, quantum
key distribution has progressed tremendously both in theory and
practice. For a recent comprehensive review, see \cite{DLB+16}.
In 1991, Ekert proposed his celebrated E91 cryptographic protocol
\cite{Eke91}, using Einstein--Podolsky--Rosen pairs (maximally entangled states), where
security was linked to the monogamous property of quantum nonlocality
\cite{Ton06}. As a result, E91 has sometimes been dubbed
`experimental metaphysics'! Interestingly, it contained the seeds
for the concept of device-independent protocols \cite{SK09}, that
would be introduced about one-and-half decades later.
Bennett's 1992 protocol, which introduced a two-state scheme, showed
that two non-orthogonal states are sufficient for quantum cryptography
\cite{Ben92}. Shor's efficient quantum algorithms for finding the
prime factors of an integer and for the discrete logarithm problem
\cite{Sho94} created a huge excitement, optimism and interest among
physicists and computer scientists because of their potential impact
on computational complexity, indicating strongly that the quantum
computers may prove to be more powerful than their classical
counterparts. The factorization algorithm, which is now known as
Shor's algorithm, got more attention because it threatened classical
cryptography, on account of its potential ability to efficiently crack
the RSA cryptographic protocol, which depends on the supposed
inability of the classical computers to factorize a large integer in
polynomial time.

The Goldenberg--Vaidman protocol \cite{GV95} shows, intriguingly,
that orthogonal states suffice for quantum key distribution. Based on a Mach--Zehnder
interferometer architecture, Goldenberg and Vaidman introduced a new paradigm in the
foundations of cryptography, where the spatial distribution of a pulse
is exploited to obviate the need for non-orthogonality of the signal
states to provide security. An experimental realization of Goldenberg--Vaidman protocol was
reported by Avella \emph{et al} \cite{ABD+10}. Later on, Goldenberg--Vaidman protocol was
generalized by various authors \cite{SPS12, YSP14, ABA+14, SP14,
 SAP14, Pat15, TSP16}, in which they established that almost all
cryptographic tasks that can be performed using a BB84 type conjugate
coding based schemes can also be performed using orthogonal state
based protocols. Specifically, they showed that it is possible to
design orthogonal state based schemes for quantum private comparison,
quantum key agreement, quantum key distribution, deterministic secure quantum communication, etc., and that thus conjugate coding is not essential for
obtaining unconditional security.

The first ever experimental demonstration of the quantum teleportation
phenomenon was reported in 1997 by Zeilinger's group at the University of
Vienna, Austria \cite{BPM+97}, who used the polarization of a photon
as a qubit. Quantum teleportation in its original form is
cryptographically insecure, but it may be used as a primitive to build
schemes for secure quantum communication.

Another new paradigm was introduced in cryptography in 1999 by Guo and Shi
who proposed a protocol based on interaction-free measurement \cite{GS99}.
In 2009, this was followed by the Noh protocol
\cite{N09}, which replaces its use of the Mach--Zehnder interferometer
with that of a Michelson interferometer. An experimental realization
of the Noh protocol was reported by Brida et al. \cite{BCD+12}.

\subsection{Intuitive security}

Quantum key distribution is intuitively secure. In BB84, Alice sends Bob a stream of states
prepared in the Pauli $X$ or $Z$ basis over an insecure channel. Bob
measures them in the $X$ or $Z$ basis randomly. Later over a
classical channel, he announces his measurement bases, and Alice
informs him which results he can keep. This step, called basis
reconciliation, creates a shared sifted key, wherein Alice and Bob
decide to assign bit value `0' to the $+1$ outcome of $X$ and $Z$,
and bit value `1' to the $-1$ of the bases. A fraction of this
sifted key is publicly announced. If Alice's and Bob's records diverge
on too many bits, they abandon the protocol run. Suppose an
eavesdropper Eve intervenes by measuring the qubits in the $X$ or $Z$
basis. At the time of key reconciliation, she knows which qubits she
measured in the right basis. Suppose Alice and Bob consume $m$ check
bits during their final test. The probability that Eve is not detected
on a given bit is $\frac{3}{4}$, or $\left(\frac{3}{4}\right)^m$ on
all $m$ bits.

A more detailed treatment of the above attack must compare Bob's and
Eve's information gain during her attack. Suppose an eavesdropper Eve
intervenes by measuring a fraction $f$ of qubits in the $X$ or $Z$
basis. She notes the result, and forwards the measured qubit. The
probability that she measures in the right basis, and thus has the
right sifted bit, is $\frac{f}{2}$. The error rate she introduces is
$e = \frac{f}{4}$, so that the mutual information between Alice and
Bob per sifted bit is \mbox{$I(A:B)=1-h(e)$}, where $I(A:B)\equiv
H(A)+H(B)-H(AB)$, and \mbox{$h(e) = -(e\log (e) + (1-e)\log (1-e))$} is Shannon
binary entropy. Eve has more information than Alice, thereby
potentially making the channel insecure \cite{CK78}, if Eve's mutual
information
\begin{equation}
I(A:E)\equiv \frac{f}{2}=2e \ge 1-h(e) \equiv I(A:B),
\end{equation} 
which happens around 17.05\%.
Here, it is assumed that Eve retrospectively knows when she measured in the right
basis. This is the case if Alice and Bob use  pseudo-random number
generators for state preparation and measurement, respectively, and
Eve is able to crack their pattern based on their public discussion
for sifting the raw key. Otherwise, Eve's information would be
$f(1-h(1/4))=4e(1-h(1/4))\le I(A{:}B)$ throughout the range $0 \le f \le 1$.

\subsection{Unconditional security}

More generally, Eve may use sophisticated attacks going beyond the
above intercept-resend method. A rigorous proof for security must be
able to cover not only general attacks on individual qubits, but also
coherent attacks on all qubits, with Eve's final manipulations
deferred until after basis reconciliation \cite{May96,May01, SP00,
 LC00}.

Here we very briefly review the proof of security of BB84 in the
spirit of \cite{SP00}. At its core are two ideas:
\begin{description} 
\item \emph{Entanglement distillation} \cite{BDS+96} via
 Calderbank--Shor--Steane (CSS) quantum error correcting codes
 or, more generally, stabilizer codes \cite{S96,CS96}. This
 corresponds to privacy amplification at the quantum level.
\item \emph{Monogamy of entanglement} \cite{CKW00}, the property that if
 Alice and Bob share singlets with high fidelity, then there is no
 third party with which Alice's or Bob's particles could be entangled
 (cf. \cite{LC00}). More generally, nonlocal no-signaling
 correlations are known to be monogamous \cite{Ton06}.
\end{description}
It is interesting that these proofs, which assume trusted devices,
make use of entanglement, which is the appropriate resource for
device-independent cryptography (discussed below), but not necessary
for security in the scenario of trusted devices. In this case,
measurement disturbance and an information vs. disturbance trade-off
suffice for guaranteeing unconditional security of key distribution,
which is proven in \cite{BBB+05}.

Regarding the first point above, viz., distillation via stabilizer
codes, an important observation is that quantum errors can be
digitized into tensor products of Pauli operators---namely bit
flips, phase flips and their products---if carefully encoded and the
errors are small enough \cite{S95,NC00}, and thereby corrected using a
classical-like (if subtler) technique.

Suppose we are given two classical linear error correcting codes $C_1
\equiv [n,k_1]$ and $C_2\equiv[n,k_2]$ such that $C_2 \subset C_1$ and
$C_1$ and $C_2^\perp$ correct up to $t$ errors on $n$ bits, with code
rates $k_1/n$ and $k_2/n$ respectively. Then, there are associated
\textit{parity check matrices} $H_1$ and $H_2$ pertaining to $C_1$ and
$C_2^\perp$, such that given a code word $w$ in a code that picks up a
bit flip error $\epsilon$ of weight of at most $t$, to become $w +
\epsilon$, it can be corrected by computing the error syndrome $H_j(w
+ \epsilon) = H_j(\epsilon)$.


The codes $C_j$ above define a $[n,k_1-k_2]$ CSS
quantum error correcting code, a subspace of $\mathbb{C}^{2^n}$. Given $u
\in C_1$, a quantum code word, which is a basis state for the quantum error correcting code, is
\begin{equation}
|u+C_2\rangle = \frac{1}{\sqrt{|\mathcal{C}_2|}}
\sum_{v \in \mathcal{C}_2} |u + v\rangle.
\label{eq:CSS}
\end{equation}
Note that $|u + C_2\rangle = |u^\prime+C_2\rangle$ if $u-u^\prime\in C_2$, so that $\ket{u+C_2}$ only depends on the coset of $C_1/C_2$
which $u$ is located in, whence the notation of (\ref{eq:CSS})
\cite{NC00}. Under $\epsilon_b$ bit flip errors and $\epsilon_f$
phase errors, the above transforms to
\begin{equation}
\frac{1}{\sqrt{|\mathcal{C}_2|}} \sum_{v \in
 \mathcal{C}_2} (-1)^{(u+v)\cdot \epsilon_f}|u + v+
\epsilon_b\rangle,
\label{eq:CSS0}
\end{equation}
 The error correcting properties of $C_1$ can be used to correct the
 $\epsilon_b$ Pauli bit flip errors by \textit{incompletely} measuring
 the quantum operators corresponding to the syndromes. After
 correcting these bit flip errors, it can be shown that applying a
 Hadamard transformation $H \equiv \frac{1}{\sqrt{2}} (X + Z)$ to each
 of the qubits, transforms these qubits to the form
\begin{equation}
\sqrt{\frac{|C_2|}{2^n}}
\sum_{w \in \mathcal{C}_2^{\perp}} (-1)^{u.w}|w + \epsilon_f\rangle,
\label{eq:CSS2}
\end{equation}
so that the phase flip errors now appear as bit flip errors, which can
be corrected using the error correcting properties of the code
$C_2^\perp$. We recover the state (\ref{eq:CSS}) after application of
$H$ to each qubit.

An application of CSS codes is to derive the Gilbert--Varshamov
bound for quantum communication, which guarantees the existence of
good quantum error correcting codes \cite{SP00}. For a $[n,k]$ CSS code correcting all errors
on at most $t\equiv\delta n$ qubits, the quantum Gilbert--Varshamov
bound says that there exist codes in the asymptotic limit such that
the code rate $k/n$ is at least $1 - 2h(2t/n)$, while giving
protection against $t$ bit errors and $t$ phase errors. Thus, in a
protocol, after correction of total errors \mbox{($\lesssim11\%$)}, Alice and
Bob share almost pure singlets hardly correlated with Eve.

The use of CSS codes for distillation can be roughly described as
follows. Suppose the channel introduces $\delta n$~errors, and Alice
and Bob encode $k$ Bell states using a $[n,k]$ quantum error correcting code correcting up to
this many errors. Alice sends Bob the qubits corresponding to the
second particle in the Bell states. Both perform identical syndrome
measurements and recovery operations on their own $n$-qubit halves of
the noisy encoded Bell pairs, recovering $k$ pairs of qubits that has
a high degree of fidelity with $k$ Bell pairs.

It is important to stress that the man in the middle
can affect quantum key distribution as much as it does classical cryptography. This
involves Eve impersonating Alice to Bob and Bob to Alice. Perhaps the
only protection for quantum key distribution against man in the middle is for Alice and Bob to share a
short inital secret (like a pass phrase) for the purpose of person
authentication. At the end of the quantum key distribution session, Alice and Bob must
store a small portion of the shared key to serve as the pass phrase
for the subsequent session. This pass phrase thus serves as a seed
that can be grown into the full key, making quantum key distribution as a kind of secret
growing protocol \cite{GRT+02}. But note that the initial seed must
have been exhanged in person or such equivalent direct means.

\subsection{Some variants \label{sec:var}}

In 2002, Bostr\"om and Felbinger introduced the Pingpong protocol
\cite{BF02} which is a two-state deterministic scheme based on quantum
dense coding. To illustrate the conceptual point that entanglement is
not required, \cite{LM05} proposed the non-entangled version of the
Pingpong protocol.

In differential phase shift quantum key distribution \cite{IWY02}, a single photon,
split into three pulses, is transmitted to Bob by Alice. Bob extracts
bit information by measuring the phase difference between two
sequential pulses by passive differential phase detection. Suitable
for fiber-based transmission, this method offers a superior key
generation rate in comparison with fiber-based BB84. The scheme has
been extended to the use of weak coherent pulses
\cite{IWY03,CAF+16}. Its security against the photon number splitting
attack \cite{IT05} and detailed security, have been studied
\cite{WTY06}. A variant of differential phase shift quantum key distribution, called the round-robin differential phase shift
protocol \cite{SYK14} has been proposed, in which a guarantee
of security is obtained even without any channel noise statistics
being monitored. The robustness of round-robin differential phase shift with regard to source flaws has
been studied \cite{MIT15}. Recently, round-robin differential phase shift has also been
experimentally realized \cite{GCL+15,LCD+16}.

The introduction of decoy states \cite{Hwa03,LMC+05,Hwa05} allows
implementation of quantum key distribution even with weak coherent pulses instead of
single-photon pulses, even in the presence of high loss. 
Kak \cite{Kak06} introduced a fully quantum scheme in which Alice and
Bob share secret bits by exchanging quantum information to and fro in
three stages, in contrast to a protocol like BB84, where classical
communication is necessary.

A research group from Toshiba Research Europe, UK, demonstrated in
2003 quantum key distribution over optical fibers about 122 km long. The commercial use of
quantum technology was initiated by this key effort \cite{GLY04}.

Building on ideas first introduced in \cite{BHK05}, in \cite{ABG+07}
quantum key distribution was analyzed under collective attacks in the device independence
scenario (discussed below), where devices are not assumed to be
trusted or well characterized.

Another direction of research in the security of quantum key distribution is to ask whether
it remains secure if only one of the two players is quantum, while the
other is classical. Boyer \emph{et al} \cite{BKM+07,BGK+09} showed that one
obtains a robust security even in this weaker situation. This is of
practical relevance, since it places a significantly lesser burden on
implementation. An open issue may be to consider how to combine
semi-quantum with device-independence (in particular, one-way
device-independence, see below).

The South Africa held 2010 Soccer World Cup marks a milestone event
for the use of quantum cryptography in a significant public event.
Quantum-based encryption was facilitated by the research team led by
F. Petruccione, Centre for Quantum Technology, University of
KwaZulu-Natal.

The use of free-space quantum communication, rather than fiber-based
optics, entered a significant phase when J.-W. Pan's group
\cite{JRY+10} implemented quantum teleportation over an optical
free-space link. Given the low atmospheric absorption under certain
wavelength ranges, this can help extend the communication distance in
comparison with a fiber link. The same research group further
reported \cite{YRL+12} the demonstration of entanglement distribution
over a free-space link of 100 km, and verifying violation of the
Clauser--Horne--Shimony--Holt inequality \cite{CHS+69}. The
high-fidelity and high-frequency techniques for data acquisition,
pointing and tracking in this process pave the way for futuristic
satellite-based quantum cryptography.

A scheme for quantum key distribution based on measurement-device independence was
proposed in \cite{LCQ12}. Its practical advantage over full device independence
is that it can tolerate the side-channel attacks and reduced
efficiency of the detectors, while doubling the secure distance using
just conventional lasers. Other works followed this: phase-encoding
for measurement-device independence \cite{TLF+12}, study of the practical aspects of measurement-device independence such as
asymmetric channel transmission and the use of decoys \cite{XCQ+13},
extending secure distance to ultra-long distances using an entangled
source in the middle \cite{XQL+13}, measurement-device-independent quantum key distribution with polarization encoding
using commercial devices acquirable off-the-shelf \cite{TLX+14}.

An experimental satellite-based quantum key distribution system, with satellite
transmitters and Earth-based (at Matera Laser Ranging Laboratory,
Italy) quantum receivers was implemented with reasonably low noise,
namely quantum bit error rate of about 4.6\%
\cite{VBD+15}. Sending quantum messages via a satellite based global
network took a further step when in 2016 China launched the \$100~million satellite mission named \emph{Quantum Experiments at Space Scale}
(QUESS) aka \emph{Micius} (after the ancient philosopher) from the
Jiuquan Satellite Launch Center. The mission aims to study the
feasibility of quantum cryptography through free-space.

\subsection{Semi-quantum protocols}

The protocols mentioned so far are completely quantum in the sense
that all the users (senders and receivers) need to be able to perform
quantum operations (like applying unitaries or measuring in
non-commuting bases) in these schemes. By a \emph{quantum user}, we mean
a user who can prepare and measure quantum states in the computational
basis as well as in one or more superposition bases (say in diagonal
basis), whose states are non-orthogonal to the computation basis
states. In contrast, a classical user is one who can perform
measurement in the computational basis only, has no quantum memory,
and who, upon receiving a qubit, can only either measure it in
computational basis or reflect it without doing anything.

An interesting question is whether all the users need to be quantum?
This important foundational question was first addressed by Boyer et
al., \cite{BKM07}, where they showed that some of the users can be
classical in a scheme called \emph{semi-quantum key distribution}.
Quite generally, such protocols, where some of the users are allowed
to be classical, are called semi-quantum. After the seminal work of
Boyer \emph{et al}, several semi-quantum schemes have been proposed
\cite{LT16, TSP16, NLW13, LQM13, KRA215, ZQZ+15}, and their security
proofs have been reported \cite{Kra15, Kra16}. For example, a
semi-quantum scheme has recently been proposed for secure direct
communication \cite{LT16}, private comparison \cite{TSP16},
information splitting \cite{NLW13}, and secret sharing \cite{LQM13}. Thus,
in brief, most of the cryptographic tasks can be done in semi-quantum
fashion, too. This is extremely important as in practical
applications end users are often expected to be classical.

\subsection{Composability}

Universal composability \cite{canetti2000universally} is a
general cryptographic framework for protocols that demands security
even when protocols are composed with other protocols or other
instances of the same protocol. For large-scale applications, clearly
composability plays an important role also in quantum cryptography
\cite{QR10}. In the context of quantum key distribution, universal composability specifies additional security
criteria that must be fulfilled in order for quantum key distribution to be composed with
other tasks to form a larger application. The ultimate goal of
security analysis would be to prove composable security against coherent
attacks. See \cite{SR08} for proofs of composable security in the
case of discrete-variable quantum key distribution and \cite{FFB+12} for continuous-variable quantum key distribution.

The universal composability model entails that a key produced via quantum key distribution is safe to be used in
other applications, such as a key to encrypt a message. Unconditional
security of quantum key distribution, as conventionally defined, does not automatically
preclude a joint attack on quantum key distribution and the message transmission based on
the resulting key. Universal composability closes this possible security loophole. As it
turns out, the conventional definition of security in quantum key distribution does entail
composable security, meaning that a key that is produced in an
unconditionally secure way is indeed safe to encode a message with
\cite{BHL+05}.

A relevant example concerns quantum key distribution being sequentially composed in order
to generate a continuous stream of secret bits. More generally, the
criteria for composability would be more stringent when mutually
mistrustful parties are involved. In this context, \cite{Unr10}
defines a universally composable security of quantum multi-party
computation. \cite{MSP+15} invokes the composability of quantum key distribution to obtain
hierarchical quantum secret sharing. A composable security has also
been defined for quantum crypto-protocols that realizes certain
classical two-party tasks \cite{FS09}. 

\section{Secure deterministic communication\label{sec:qsdc}}

There are several facets of secure quantum communication, which can in
principle be derived by composing quantum key distribution and having access to a secure
random number generator. In this subsection we aim to provide an
interconnection between them \cite{TPB16, P13} via specific examples.

To begin with we describe a scheme for controlled quantum dialogue.
There are three users Alice, Bob and Charlie, such that the
communication channel between Alice and Bob is supervised by Charlie,
who is referred to as controller. Alice and Bob both can send
classical information to each other in a secure manner using this
quantum channel, which constitutes a quantum dialogue. However,
Charlie fully determines whether the channel is available to them
both. Further, a requirement of quantum dialogue is that classical communication
between Alice and Bob should be transmitted through the same quantum
channel and that it should be transmitted simultaneously (namely, there
must be a time interval, during which the information of both parties
would be in an encoded state in the same channel).

Here, it is important to note that Alice and Bob need to be
semi-honest (a semi-honest user strictly follows the protocol, but
tries to cheat and/or obtain additional information remaining within
the protocol), as otherwise they may create an independent quantum
channel of their own and circumvent the control of Charlie. Now, we
may briefly describe a simple scheme of controlled quantum dialogue as follows \cite{TP15}:
\begin{description}
\item \emph{Step~1}: Charlie prepares $n$ copies of a Bell state, diving
 them into two $n$-qubit sequences $S_{A}$ and $S_{B}$, with the
 first and second halves of the Bell pair, respectively. Then, he
 transmits both $S_{A}$ and $S_{B}$ to Bob, after suitably permuting
 $S_{B}$. It is assumed that all qubit transmissions are secure, with
 the possible inclusion of decoy qubits, which are inserted to test
 for an eavesdropper and dropped afterwards \cite{STP+16}.
\item \emph{Step~2}: Using Pauli operations in the manner of quantum
 dense coding \cite{NC00} (whereby $I$, $X,$ $iY$, and $Z$ correspond
 to encoded bit values 00, 01, 10, and 11, respectively), Bob encodes
 his message in the qubit string $S_{A}$, which he then transmits to
 Alice.
\item \emph{Step~3}: After using the same method to encode her secret
 message, Alice transmits back the sequence $S_A$ to Bob.
\item \emph{Step~4}: Charlie reveals the permutation used. On this
 basis, Bob pairs up the partner particles and measures them in the
 Bell basis.
\item \emph{Step~5}: Bob publicly announces the outcomes of his
 measurements, which allows each party to extract the other's message
 using knowledge of her/his own encoding and that of initial Bell
 state Charlie prepared.
\end{description}

Without Charlie revealing the particular permutation used, semi-honest
Alice and Bob cannot decode the other's message, thereby ensuring
Charlie's control. Moreover, just before step 4, both Alice and Bob's
messages are encoded at the same time in the channel, which ensures
satisfaction of the quantum dialogue requirement.

Charlie's choice of Bell state, if publicly known, would lead to
information leakage, which is often considered to be an inherent
feature of quantum dialogue and variants thereof. This problem can be eliminated if
Charlie chooses his Bell state randomly, informing Alice and Bob of
his choice via quantum secure direct communication or deterministic secure quantum communication \cite{BST+16}.

The above scheme can be turned into other crypto-tasks. If Bob,
instead of Charlie, prepares the Bell states initially (with the
difference of Charlie's announcement being absent in step 4), then the
above scheme reduces to quantum dialogue, of the type introduced at first by Ba An
\cite{AN05}. This is called the Ba An protocol for quantum dialogue.

Likewise, a quantum dialogue scheme can always be obtained from a controlled quantum dialogue scheme.
Further, in a quantum dialogue scheme, restricting one of the players, e.g., Alice,
to trivial encoding (namely, simply applying Pauli $I$ operation), we
obtain a protocol for quantum secure direct communication,
whereby Bob can communicate a message to Alice without the prior
distribution of a key. In this way, any quantum dialogue can be turned into that for
quantum secure direct communication. In quantum secure direct communication, a meaningful message is typically sent by the sender.
Instead, transmission of a random key turns quantum secure direct communication into a quantum key distribution.
Therefore, any quantum secure direct communication protocol can be turned into a quantum key distribution protocol
\cite{TPB16}.

Likewise, suppose that in a quantum dialogue scheme, Alice (resp., Bob) transmits
key $k_{A}$ (resp., $k_{B})$ to Bob (resp., Alice), after which they
adopt $K=k_A\oplus k_B$ as the secret key for future communication,
this constitutes a protocol for quantum key agreement, in which
each player contributes equally to $K$, such that each bit of $K$
cannot be unilaterally determined by either player. In this way, a
quantum key agreement scheme can always be obtained from that for quantum dialogue. Also, in
asymmetric quantum dialogue \cite{BST+16}, a special case of the quantum dialogue scheme,
involves Alice and Bob encoding an unequal amount of information (say,
Alice sending $m$ bits, and Bob sending $2m$ bits).

Other types of reduction are possible. In the above scheme for controlled quantum dialogue,
suppose Charlie retains sequence $S_{B}$ and only transmits $S_{A}$
securely to Alice, who encodes her secret message using the dense
coding method and then transmits the resultant qubit string to Bob.
Upon reception, Bob encodes his secret using the same rule and sends
the resultant sequence to Charlie, who finally measures each received
particle with its partner particle retained by him, in the Bell basis.
If in each case, he obtains the original Bell state, the Alice's and
Bob's secrets are identical. This follows simply from the fact that
$I=XX=(iY)(iY)=ZZ=I_{2}I_{2}$, ensuring that two encoded messages are
identical, then the travel qubits return as they left.

Therefore, a quantum dialogue or controlled quantum dialogue scheme can always be turned into one for
quantum private comparison, which allows a third party to
compare the secrets of two parties without being able to know their
secrets \cite{TSP16}. This quantum private comparison is suitable for the socialist
millionaire problem or Tierce problem \cite{FST01}, which is a secure
two-party computation requiring two millionaires to find out if they
are equally rich, without revealing how rich each is (unless of course
they are equally rich). In brief, a modification of quantum dialogue or controlled quantum dialogue
provides a solution for quantum private comparison, the socialist millionaire problem and a
few other related problems.

Just as a quantum dialogue protocol can be turned into a quantum secure direct communication one, a controlled quantum dialogue protocol can
be turned into one for controlled quantum secure direct communication (technically, actually one for controlled deterministic secure quantum communication).
Now, controlled deterministic secure quantum communication can be used in a quantum e-commerce situation, where Charlie
represents a bank, Alice a buyer and Bob an online shop. To make a
purchase, Alice intimates Charlie, who executes step 1 above. Next,
Alice encodes her purchase information in $S_{A}$, which she sends to
Bob, who in turn informs Charlie of having received an order worth a
specific amount from a certain buyer, whose identity is verified by
Charlie, who then reveals the relevant permutation operation. Bob
then performs Bell measurement and knows about Alice's order.
Therefore, a quantum e-commerce protocol of this type is really a
straightforward modification of a controlled quantum secure direct communication scheme. 

In fact, in the recent past several schemes of quantum \mbox{e-commerce} and
other similar applications of quantum cryptography have been proposed
by various groups \cite{WCF13,HYJ15}, that have established
that quantum cryptography can be used for various practical tasks
beyond key distribution and secure direct communication. Specifically,
sealed-bid auctions \cite{ZNZ10, Nas09,YNW09} and other variants of
auctioning (e.g., English and Dutch auctions) can be perfromed using
quantum resources \cite{PS03,PS08}. Binary voting can also be
performed using quantum resources (cf. \cite{TSP17} and references
therein).

\section{Counterfactual quantum cryptography\label{sec:CQC}}

Counterfactual quantum communication transmits information using the
non-travel of a photon between Alice and Bob \cite{Joz98, MJ01,
 HRB+06, Vai07}. It is based on interaction-free measurement
\cite{EV93, GS99, KWH+95}, where the presence of an object is detected
without directly interrogating it. Famously known as the
Elitzur--Vaidman scheme for bomb detection, it involves photon
interferometery used to ascertain the presence of a quantum object in
one of the arms without the photon actually passing through it. The
single-photon injected into the beamspliiter of this set-up always
exits one particular output port labelled as the bright port. The
presence of an object in one of the arms of the interferometer permits
the single photon to exit not from the bright port, but through the
port that is otherwise dark.

Experimental realizations proved that indeed such interaction-free
measurements are possible \cite{KWH+95}. Further, a proposal to
enhance its efficiency towards 100\% using chained unbalanced
beamsplitters, wherein repeated measurements of the initial state in
order to arrest evolution, simulating the quantum Zeno effect, was put
forth. The scheme works as follows: A single-photon incident on a
beamspliiter after $M$ cycles exits from the bright port but the
presence of a detector in these ports restricts the photon to be
always in the lower arm and exit from the dark port. The chained
action leads to the evolution:
\begin{align}
|\texttt{block}\rangle|0\rangle &\rightarrow \cos^{M}{\eta}
|\texttt{block}\rangle|0\rangle \nonumber \\
|\texttt{pass}\rangle|1\rangle &\rightarrow (\cos{(M\eta)}|0\rangle
+\sin{(M\eta)}|1\rangle)
\end{align}
where $M$ is the number of interferometric cycles, and the first
equation indicates absorption at the obstacle. In 2009, Noh proposed
the well-known counterfactual quantum protocol for cryptography
\cite{N09}.

Though counterfactual quantum cryptography may not be so useful for
long-distance communication, it is interesting conceptually
\cite{SSS1,SS2}. Schemes to improve the efficiency of counterfactual
quantum key distribution protocols \cite {SW10, ZLZ14}, security analysis of such schemes
under various attacks such as intercept-resend and counterfactual
attacks\cite{YLC+10,ZWT12,ZWT+12}, experimental realisation using
different set-ups \cite{BCD+12,LJL+12,YLY+12}, direct communication
protocols \cite{SSS14} and counterfactual generation and distribution
of entanglement \cite{SS15} have contributed towards better
understanding of applying counterfactuality. The basic idea of the
direct communication protocol is to ensure counterfactual transfer of
information using the chained beamsplitter approach mentioned earlier.
$M$-chained unbalanced beamsplitters are nested within $N$-chained
outer unbalanced beamsplitters. By suitably choosing $M$ and $N$, one
can achieve direct communication between Alice and Bob. It has been
further argued that this is fully counterfactual \cite{SLA+13}, an
interpretation that has been debated. For an alternative perspective,
see \cite{Vai14,Vaid14}, but also \cite{SLA+14,Gis13}. Recently, the
proposal in \cite{SLA+13} for direct counterfactual communication has
been implemented experimentally \cite{cao2017direct}.

By letting the obstacle to be in a superposition state, as follows:
\begin{align}
(\alpha|\texttt{block}\rangle&+\beta|\texttt{pass}\rangle)|0\rangle 
\rightarrow \alpha \cos^{M}{\eta}|\texttt{block}\rangle|0\rangle \nonumber \\
&+ \beta |\texttt{pass}\rangle (\cos{(M\eta)}|0\rangle+\sin{(M\eta)}|1\rangle) \nonumber \\
&\rightarrow \alpha |\texttt{block}\rangle |0\rangle + \beta 
|\texttt{pass}\rangle |1\rangle 
\end{align}
An idea along this line can be used to counterfactually transmit a
qubit, as against a bit \cite{GCC2+14, GCC+14, S14a}.

\begin{figure*}[t!]
\begin{center}
\includegraphics[width=130mm]{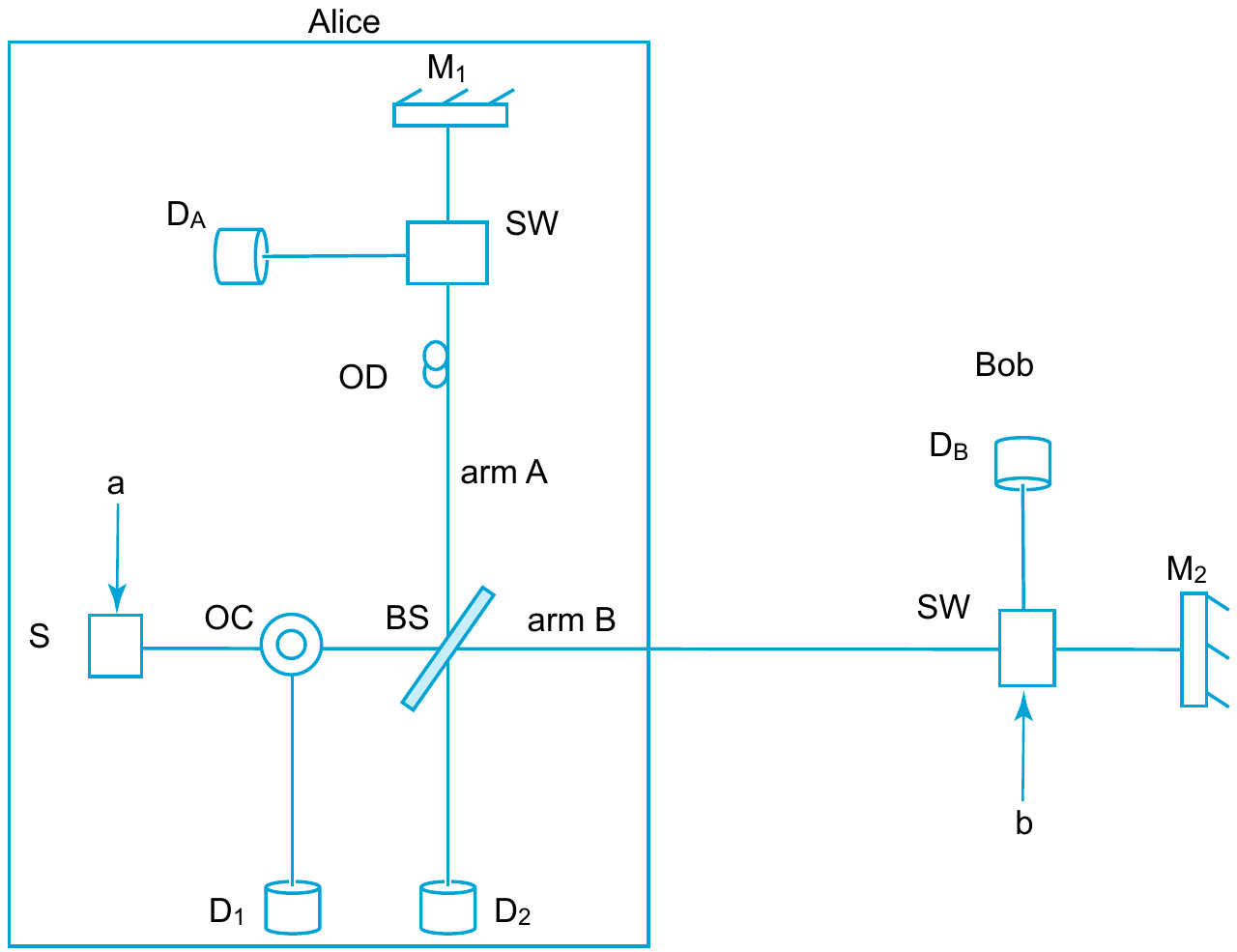}
\end{center}
\caption{\color[HTML]{000000} Basic set-up for counterfactual \cite{N09} or semi-counterfactual \cite{SSS1} quantum key distribution based on a Mach--Zehnder interferometer. Alice's station consisting of the source (S)
initiates the protocol by sending light pulses through the optical
circulator (OC) to the beamsplitter (BS), which splits
them into beams along arms A and B. The optical delay (OD)
maintains the phase between the arm by compensating for the
path difference in the two arms. Light along arm A is subjected
to absorption by detector $D_A$ or reflection from mirror $M_1$ by Alice based on her switch (SW)
state. Likewise by Bob along arm B who also possesses a switch (SW), a detector $D_B$
and a mirror $M_2$.}
\label{fig:cfa}
\end{figure*}

The well-known counterfactual protocol Noh09 \cite{N09} is briefly
explained here. Alice and Bob are connected to each other through one
of the arms of a Michelson interferometer (arm B). The other arm A
is internal to Alice's station and is inaccessible to the outside
world. A photon traveling along arm A is always reflected using a
Faraday mirror ($M_1$). In addition, Alice is also equipped with a
single-photon source which prepares polarization states in the
vertical ($V$) or horizontal ($H$) direction, based on the output of a
quantum random number generator ($a$). Bob's station also consists a
quantum random number generator ($b$) whose output decides whether a
reflection using Faraday mirror ($M_2$) or a detection using a detector
$D_B$ is to be applied. $R_B$ controls a switch $Q$ whose polarization
state $P$ (pass $V$ and block $H$) or $B$ (block $V$ and pass $H$)
determines which of the above operations is to be applied. The
protocol is as follows (cf. Figure \ref{fig:cfa}):

\begin{enumerate}
\item Alice prepares polarization states randomly in $V$ or $H$ states
 based on $a$ and transmits it to Bob.
\item Bob applies $P$ or $B$ randomly based on $b$. The following
 table gives the conditional probabilities based on Alice and Bob's
 joint action:
\begin{center}
\begin{tabular}{|c|c|c|c|}	
\hline		
(Alice,Bob) & $D_1$ & $D_2$ & $D_B$ \\
\hline 
$(V,P)$ or $(H,B)$ & $0$ & $1$ & $0$ \\
\hline
$(V,B)$ or $(H,P)$ & $RT$ & $R^2$ & $T$ \\
\hline
\end{tabular}
\end{center}
$D_1$ and $D_2$ are detectors in Alice's station. $R$ and $T$ are the
coefficient of reflectance and transmittance of the beamsplitter
respectively such that $R+T=1$.
\item At the end, $D_1$ detections lead to the generation of secret
 key and $D_2$ detections are used for detecting eavesdropping. $D_1$
 detection is counterfactual in the sense that the photon did not
 travel to Bob and his blocking action leads to a remote detection by
 Alice. In some sense, the photon takes into account Bob's choices
 before detection.							
\end{enumerate}
Future directions here could explore applying the counterfactual
paradigm to other crypto-tasks besides quantum key distribution.

\section{Device independent scenarios\label{sec:DIS}}

We already noted that a classical cryptographic protocol is secure
only under some assumptions about the hardness of performing some
computational tasks. In contrast, BB84, B92 and other protocols for
quantum key distribution, mentioned above, are unconditionally secure in the sense that
their security proof is not based on such computational assumptions,
but instead is guaranteed by quantum physical laws.

However, the security proofs assume that the practical realization is
faithful to the theoretical specifications, and that the devices used
by Alice and Bob are trusted and that the duo have perfect control
over the devices used for the preparation of the quantum states,
manipulation and their measurement. Devices are also assumed to be
free from any side channels that would leak secret information from
the laboratories of Alice and Bob.

\subsection{Side-channel attacks}

Quantum key-distribution promises unconditional security under the
assumption of perfect implementation of the protocols in the
real-world. But, imperfections in the experimental set-up creates
side-channels that can be employed by a malicious eavesdropper to
compromise the security without the knowledge of the legitimate
participants Alice and Bob. Side-channel attack allows Eve to gain
information based on certain behavioural patterns of the devices used
for key-distribution and does not depend upon the theoretical security
\cite{BP12}.

Some examples of side-channels are detector clicks, dark counts and
recovery time of the detectors, electromagnetic leaks. Sometimes the
side-channel attacks are so powerful that the basis information may be
leaked and render the protocol completely insecure. Such side-channel
attacks identified \cite{NFM+09,RVC+15} the danger posed by not being
able to completely characterise sources and detectors, leading to the
device-independent paradigm \cite{BKH+16}.

Here, we list some powerful quantum hacking attacks and
countermeasures on commercial quantum key distribution systems:
\begin{itemize}
\item Time-shift attacks, which make use of the detection efficiency
 loophole, which plays a key role in the Bell inequality tests
 \cite{ZFQ+08}. Here we may count bright illumination pulses to
 control single-photon avalanche photodetectors \cite{LWW+10},
 passive detector side channel attacks \cite{L15} and detector
 blinding attacks. In particular, information leakage due to the
 secondary photo-emission of a single photon avalanche detector can
 be countered by backflash light characterization \cite{MDT+16}.
\item Time side channel attack where the timing information revealed
 during the public communication during Alice and Bob is used to
 extract some parts of the secret key \cite{LK07}.
\item Optical side-channel attacks in order to gain information about
 the internal states being propagated in the channel \cite{FH08}.
\item Source attacks based on tampering of the photon sources in the
 measurement device-independent paradigm \cite{SXJ+15}.
\item Preventing side-channel attacks in continuous-variable quantum key distribution by modulating the
 source signal appropriately to compensate for a lossy channel
 \cite{DUF16}.
\end{itemize}

\subsection{Device-independence}

The practical realization of an otherwise unconditionally secure quantum key distribution
protocol will involve the use of untrusted devices \cite{MHH+97},
whose imperfections may be exploited by a malicious eavesdropper to
undermine its security. In 1998, Mayers and Yao \cite{MY98}
introduced the concept of quantum cryptography with the guarantee of
security based only on the passing by the measurement data of certain
statistical tests, under the assumptions of validity of quantum mechanics and the
physical isolation (no information leakage) of Alice's and Bob's
devices; in other words, a quantum key distribution set-up that can self-test.

In \cite{BHK05}, it was shown how a single random bit can be securely
distributed in the presence of a non-signalling---not just quantum---eavesdropper. This qualitative argument was made quantitative by
several following works, providing efficient protocols against
individual attacks \cite{AMP06, SGB+06}, and subsequently collective
attacks \cite{MRC+09,Mas09,HRW10} against a non-signaling
eavesdropper. Better key rates, but assuming an eavesdropper
constrained by quantum laws, are reported in \cite{ABG+07, PAB+09,
 MPA11}. All these proofs of security require an independence
assumption, namely that successive measurements performed on the
devices commute with each other \cite{MPA11}. While \cite{BCK12a}
fixes this issue, by allowing Alice and Bob to use just one device
each, it is inefficient and lacks noise tolerance. The device-independent protocol of
\cite{VV12} reports an improvement guaranteeing the generation of a
linear key rate even with devices subject to a constant noise rate,
but relaxing other assumptions such as the availability of several
independent pairs of devices.

We briefly mention the connection of device independence and
nonlocality. A necessary condition in order to guarantee security in
the scenario where devices are not assumed to be trustworthy---characteristic of the device-independent scenario---is that Alice's and Bob's joint
correlation $P(x,y|a,b)$, where $a,b$ are the respective inputs and
$x,y$ their respective outputs, must be such that
\begin{equation}
P(x,y|a,b) \ne \sum_\lambda P(x|a,\lambda)P(y|b,\lambda)P_\lambda,
\label{eq:factorizable}
\end{equation}
where $P(x|a,\lambda)$ and $P(y|b,\lambda)$ are arbitrary probability
distributions for Alice and Bob; and $P_\lambda$ is the probability
distribution of some underlying variable $\lambda$. For if this were
not so, then in principle, knowing $\lambda$, Eve would be able to
determine the outcomes of Alice and Bob, when they announce $a$ and
$b$ publicly during the key reconciliation step. This entails that
$P(x,y|a,b)$ must be nonlocal, namely it should violate a Bell-type
inequality, making the sharing of quantum entanglement necessary.

Other than this, the quantum apparatuses used by Alice and Bob are
viewed as black boxes, with no assumption made about the internal
workings. Interestingly, the root concept for device-independent quantum key distribution was implicit as
early as 1991 in the E91 protocol \cite{Eke91}, but its true
significance was not recognized before the advent of the study into device-independent
cryptography. Because the security of any device-independent scheme requires nonlocal
correlations, which is in practice an expensive and delicate resource,
it would be difficult to achieve full device independence. For
example, the detector efficiencies are usually too low to support full device-independent
security. Although the hope for practical realization of device-independent quantum key distribution has
been raised by recent loophole-free Bell experiments
\cite{HBD+15,SMC+15,Giu15}, the secure key rates are expected to be
quite low even for short distances.

Although we have generally talked of quantum key distribution in the context of device independence,
other tasks can also be considered in this framework, among them
self-testing of the $W$ state \cite{WCY+14} $$W\equiv\frac{1}{\sqrt{3}}(|001\rangle +
|010\rangle + |100\rangle)$$ and that of any two
projective observables of a qubit \cite{kan17}, have been reported.

Of interest here is the device-independent quantum key distribution protocol based on a local Bell test
\cite{LPT+0}. Several relaxed variants of device-independent quantum key distribution idea (including
semi-device-independent and one-way device-independent) have been proposed \cite{LCQ12, BCW+12,
 ZBZ+15, PB11*, Woo16} and are briefly discussed below.
 
\subsection{Measurement-device independence \label{sec:mdi}}

A more feasible solution than device-independent quantum key distribution is the
measurement-device-independent quantum key distribution
\cite{LCQ12} scheme, which builds on
\cite{BHM96,Ina02}. Using weak coherent light pulses along with decoy
states, the measurement-device-independent quantum key distribution protocol is made immune to all side-channel
attacks on the measurement device, often the most vulnerable part.
However, it is assumed in  measurement-device-independent quantum key distribution that Eve cannot access state
preparation by Alice and Bob. The security of measurement-device-independent quantum key distribution against general
coherent attacks, exploiting the effect of finite data size, has been
proven in \cite{Cur14}. In this context, see \cite{BP12}, which
proposes using quantum memory and entanglement to replace all real
channels in a quantum key distribution protocol with side-channel-free virtual
counterparts.

Measurement-device-independent quantum key distribution, in contrast to a full device-independent scheme, requires neither almost perfect
detectors nor a qubit amplifier nor a measurement of photon number
using quantum non-demolition measurement techniques
\cite{LCQ12,XCQ+15}; also cf. related references cited in Section
\ref{sec:QKD}. The most recent developments of the measurement-device-independent quantum key distribution
scenario, including its strengths, assumptions and weaknesses are
reviewed in \cite{XCQ+15}.

The basic idea behind measurement-device independence is that Alice and Bob transmit weak coherent
pulses representing randomized BB84 polarization states to a central
untrusted Bell state measurement station, manned by Charlie or even
Eve. The probabilistic production of Bell states can be shown to lead
to a secure bits, even if the untrusted station uses only linear
optics.

Measurement-device-independent schemes have been experimentally realized by various groups
\cite{TLX+14, LCW+13, TYC+14}. It has even been demonstrated through
a distance over 200 km, whereas a full device-independent scheme is yet to be realized
experimentally. For discrete-variable measurement-device-independent quantum key distribution, the key rate for
practical distances turns out to be just 2 orders of magnitude below
the Takeoka--Guha--Wilde bound \cite{TGW14}, enabling this method to meet the high
speed demand in metropolitan networks.

\subsection{Detector-device-independence \label{sec:ddi}}

Whereas the key rate of measurement-device-independent quantum key distribution scales linearly with transmittance of
the channel (just as with conventional quantum key distribution), it has the drawback that
its key rate scales quadratically (rather than linearly, as in
conventional quantum key distribution) with detector efficiency \cite{DLB+16}, which can be a
practical problem if detectors of sufficiently high efficiency are not
available. Detector-device-independent quantum key distribution aims to combine the
efficiency of the conventional quantum key distribution protocols with the security of
measurement-device-independent quantum key distribution \cite{GRF+15,Lim14}. In detector-device-independent quantum key distribution, receiver Bob decodes photon
information from an insecure channel using a trusted linear optics,
followed by a Bell state measurement with untrusted detectors.

The advantage of detector-device-independent quantum key distribution over measurement-device-independent quantum key distribution is that key rate scales linearly
(rather than quadratically) with detector efficiency, essentially
because it replaces the two-photon Bell state measurement scheme of measurement-device-independent quantum key distribution with a
single-photon Bell state measurement scheme \cite{Kim03}. (In a single-photon Bell
state, spatial and polarization modes---each representing a bit---are
entangled.) However, the security of detector-device-independent quantum key distribution against all detector
side-challels remains yet to be shown. It is known \cite{Qi15} that
either countermeasures to certain Trojan horse attacks \cite{GFK+06}
or some trustworthiness to the Bell state measurement device is required to guarantee the
security of detector-device-independent quantum key distribution (as against the strong security of measurement-device-independent quantum key distribution, where
such assumptions are not needed.) Indeed, a simple implementation of
a detector-device-independent quantum key distribution protocol can be built directly on the standard
phase-encoding-based BB84 quantum key distribution \cite{LLY+15}. 

\pagebreak
\subsection{One-sided device-independence \label{sec:1sdi}}

Further, violation of Bell's inequality or equivalently the use of a
Bell nonlocal state can ensure the security of a device-independent quantum key distribution mentioned
above, but if one of the users (Alice or Bob) trusts her/his
devices then we obtain a weakening of device-independent quantum key distribution, known as one-sided device-independent quantum key distribution \cite{BCW+12}, whose security does not require the
violation of Bell's inequality, but rather a weaker type of
nonlocality, namely quantum steerability.

The condition (\ref{eq:factorizable}) is symmetric between Alice and
Bob. Now suppose $P(x,y|a,b)$ satisfies the asymmetric but weaker
condition \cite{WJD07}:
\begin{equation}
P(x,y|a,b) \ne \sum_\lambda P(x|a,\lambda)P^Q(y|b,\lambda)P_\lambda,
\label{eq:steerable}
\end{equation}
where $P^Q(y|b,\lambda)$ is any quantumly realizable probability
distribution for Bob. Such a state is said to be steerable, and can be
pointed out by the violation of a steering inequality. Steering is a
stronger condition than nonseparability, but weaker than nonlocality.

C.~Branciard \emph{et al} \cite{BCW+12} first studied the security and practicability of
one-sided device-independent quantum key distribution, which belongs to a scenario intermediate between
device-independent quantum key distribution and standard quantum key distribution. This makes it more applicable to practical
situations than the latter. Just as a sufficient violation of a
Bell-type inequality is necessary to establish device-independent quantum key distribution, so a
demonstration of steering is necessary for security in the one-sided device-independent quantum key distribution
scenario.

It may be noted that the prepare-and-measure schemes of quantum key distribution that do
not use entangled states (e.g., BB84 and B92 protocols) can also be
turned into entanglement-based equivalents, from which we can obtain
their device-independent counterparts by employing suitable Bell-type inequalities.
For example, M.~Lucamarini \emph{et al} \cite{LVG+12} presented a device-independent version of a modified B92
protocol. T.~Gehring \emph{et al} \cite{Geh15} reported an experimental
realization of continuous-variable quantum key distribution with composable and one-sided device-independent
security against coherent attacks. A one-sided device-independent implementation of
continuous-variable quantum key distribution has been experimentally implemented, wherein the key rate is
directly linked to the violation of the Einstein--Podolsky--Rosen steering inequality
relevant to the context \cite{walk2016experimental}.

Here it would be apt to note that for pure states, entanglement,
steering and nonlocality are equivalent. However, for mixed states
they are different and all Bell nonlocal states are steerable, and all
steerable states are entangled, but not the other way in each case,
namely entanglement is the weakest and nonlocality the strongest
nonclassicality condition among these.

\subsection{Semi-device independence}

In quantum mechanics, an entangled measurement is represented by an operator, at
least one of whose eigenstates corresponds to an entangled state. In
the \textit{semi-device-independent} approach \cite{VN11},
one can certify that the measurement is indeed entangled on basis of
the measurement statistics alone, provided it can be assumed that the
states prepared for testing the measurement apparatus are of fixed
Hilbert space dimension, even if uncharacterized otherwise. This
approach has been applied to other quantum information processing
tasks, among them cryptography \cite{PB11*}.

Now, it is possible to test the dimension of a physical system in a device-independent
manner, namely on basis of measurement outcomes alone, without requiring
the devices to be characterized, by means of Bell inequalities
\cite{BPA+08,NTV14} or bounds pertaining to quantum random access
codes \cite{WCD08}. More recently, the semi-device-independent approach has been
applied to estimate classical and quantum dimensions for systems in a
prepare-and-measure setup \cite{GBH+10,BNV13}. Experimental
realization of these ideas have been reported \cite{HGM+12,ABC+12}, as
well as their application to cryptography \cite{PB11*} and random
number generation \cite{LPY+12,LYW+12}.

For prepare-and-measure protocols in quantum information processing,
since quantum nonlocality is out of question, a more natural notion of
device independence applicable is the semi-device-independent scenario. This uses the
notion of bounding the classical or quantum dimension required to
reproduce the observed quantum correlations by measurements on
transmitted particles prepared in specific states \cite{GBH+10,BQB14}.
Let
\begin{equation}
P(y|\alpha,b) = \textrm{Tr}(\rho_\alpha \Pi^y_b)
\label{eq:mez}
\end{equation}
denote Bob's probability for getting outcome $y$ given measurement $b$
acting on state $\rho_\alpha$ transmitted by Alice, with $\Pi^y_b$
being the corresponding quantum mechanical measurement operator. 

A dimension witness for the prepare-and-measure scenario has the form
\begin{equation}
\sum_{\alpha,b,y} f_{\alpha,b,y} P(y|\alpha,b) \le C_d,
\label{eq:semiDI}
\end{equation}
where $f_{\alpha,b,y}$ are a set of real numbers and $C_d$ is a
positive real number. Violation of (\ref{eq:semiDI}) would mean that
no classical particle of dimension $d$ could have generated the
observed experimental correlation $P(y|\alpha,b)$. More generally,
one can bound the quantum dimension, also \cite{GBH+10}. This
violation serves as the basis for semi-device-independent security, just as violation
of a Bell inequality serves as the basis for device-independent security.

In \cite{Woo16} a semi-device-independent version of BB84 protocol has been
presented using the notion of semi-device-independence introduced in \cite{PB11*}.
Similar device-independent and semi-device-independent generalizations of other protocols are also
possible, and a general prescription for the same is a problem worth
exploring.

\subsection{Security in a post-quantum world}

There is an intrinsic and quite reasonable assumption in the security
proof of all the above protocols on the validity of quantum mechanics.
What would happen to the keys if the nature is found to obey a theory
other than quantum mechanics? It turns out that so long as a theory
admits a no-cloning theorem, then (possibly assuming trusted devices)
security is possible \cite{ASA16}, whereas device-independent security would be
possible if it is a nonlocal non-signaling theory. In fact, the
concept of device independence can be adapted to provide security
against even a post-quantum Eve constrained, assuming only the
no-signaling principle \cite{BHK05,AGM06}.

\section{Further applications of quantum cryptography\label{sec:post}}

We shall now survey various crypto-tasks other than quantum key distribution for which
quantum cryptographic schemes have been proposed.

\subsection{Quantum random number generation}

Apart from key distribution, current levels of quantum technology
suffice for providing a good source of genuine randomness, which is
important for cryptography and in algorithms for simulation. As noted
above, quantum random number generators are available
commercially now. By \emph{genuinely random} we refer to a source
whose output is unpredictable and irreproducible according to known
physical laws. This stands in contrast to
pseudo-random number generator, which generates strings which are pre-determined
according to a deterministic algorithm. One may then hope that the
numbers produced by a pseudo-random number generator are distributed indistinguishably from a
uniform probability distribution. The robustness of pseudo-random number generators is an issue
that would merit careful consideration \cite{DPR+13}.

From the perspective of algorithmic information theory, genuinely
random strings are incompressible, namely their Kolmogorov complexity
is not smaller than the string's length \cite{Cha87}, whereas
pseudo-randomness is algorithmically compressible. Kolmogorov
complexity of string $S$ refers to the length in bits of the shortest
computer program (in a given programming language) that generates $S$
as its output. However, in general, randomness cannot be proven
because Kolmogorov complexity is uncomputable. For practical purposes,
the randomness of given data may be evaluated by running the data
through standard statistical tests for random number generators, such
as the suite provided by the National Institute for Standards and
Testing \cite{NIST}.

The most straightforward quantum random number generator exploits the randomness of outcomes in
quantum measurements, for example, by reading off the output of a
50/50 beam splitter \cite{ROT94,SGG+00,JAW+00}. Other sources of
randomness include single photon arrival times
\cite{DYS+08,WLB+11,NZZ+14,SR07}, a laser's phase noise
\cite{QCL+10,UAI+08,AAJ+14} and vacuum fluctuations
\cite{GWS+10,SAL11}. Mobile phone cameras provide a good, if
classical, source of randomness \cite{SMZ+14}. 

An important issue here is to estimate the entropy of the randomness
source, namely the raw random bits generated, from which truly random
bits can be extracted \cite{NT99}. Sophisticated techniques have been
developed to estimate entropy in specific cases
\cite{FRT13,MXX+13}. However, these methods are somewhat difficult to
implement and do not easily lend themselves to generalization nor to
easy real-time monitoring. Device-independent quantum random number generator provides a
possible solution \cite{Col09,PAM+10}, which makes use of suitable
tests of violation of a Bell-type inequality \cite{PAM+10,CMA+13},
making them however not so simple to implement in practice as a basis
for a quantum random number generator.

Semi-device-independent certification of randomness \cite{LYW+12} is
simpler, but not entirely free from loopholes in practice
\cite{DPG+15}. A method based on the uncertainty principle, but
requiring a fully characterized measurement device, has recently been
proposed \cite{VMT+14}. As an improvement, Lunghi \emph{et al}
\cite{LBL+15} have proposed a self-testing prepare-and-measure quantum random number generator
protocol based on measuring a pair of incompatible quantum
observables.

The incompatibility, and consequently the amount of genuine
randomness, can be quantified directly from the experimental
data. These authors have also reported a practical implementation
using a single-photon source and fiber optical communication channel,
through which they report a 23-bit throughput of genuine randomness at
99\% confidence level \cite{LBL+15} .

\subsection{Quantum secret sharing\label{sec:QSS}}

Secret sharing is a crypto-task where a dealer splits a secret into
two or more shares and distributes them among multiple agents such
that only designated groups of agents (who define an access structure)
can reconstruct the full secret.

Secret sharing is a cryptographic primitive used widely to design
schemes for digital signature, key management, secure multiparty
computation, etc. Classical secret sharing, first proposed
independently by Shamir and Blakely, makes certain computational
assumptions about complexity, making its security computational rather
than unconditional. Quantum mechanics has given grounds for hope here
\cite{HBB99,KKM+99}. The original proposal for quantum secret sharing \cite{HBB99} distributes a 3-qubit state among three
participants:
\begin{eqnarray}
\hspace*{-6mm}|\psi\rangle &=& \frac{1}{\sqrt{2}}(|000\rangle+|111\rangle)\nonumber\\
&=&\frac{1}{2}(|{+++}\rangle+|{--+}\rangle+|{-+-}\rangle+|{+--}\rangle)
\label{eq:qss}
\end{eqnarray}
The three parties measure in $X$ or $Z$ basis. On a quarter of the
time (that may be established by classical communication), all three
would have been measured in the same basis, and it is clear from Eq.
(\ref{eq:qss}) that Bob and Charlie can reconstruct Alice's bit
(designated the secret) by combining their results. Any attempt by a
third party to find out the secret disrupts the correlation, which can
be detected by the legitimate parties by announcing full outcomes in
some trial runs. Another important aspect to be considered is that
one or more of the participants themselves could be colluding to
cheat. A full proof of security must also consider such player
collusion scenarios.

An extension of the above is a quantum $(N,k)$ threshold scheme,
where a quantum secret, split among $N$ parties, can be reconstructed
only if at least $k$ parties combine their shares. The no-cloning
theorem implies that $2k > N$. Threshold schemes have similarities
with quantum error correcting codes \cite{CGL99}. Generalizations of
quantum secret sharing to more general access structures
\cite{G00,SS05} and the use of various relatively easily prepared
quantum states beyond the Greenberger--Horne--Zeilinger states \cite{MP08a} have been studied,
as well as their use in the related task of quantum information
splitting \cite{MP08b,KFM+10}.

The concept of quantum secret sharing has been further generalized in various ways, among
them: hierarchical quantum secret sharing \cite{SP13},
hierachical dynamic quantum secret sharing \cite{MSP+15}. Further, in a recent
direction \cite{THZ+15}, quantum secret sharing based on a $d$-level particle (with $d$
being an odd-prime in order to exploit the cyclic property of the
$d+1$ mutually unbiased bases in these dimensions), rather than
entanglement, has been studied. Suppose the vector in this system is
denoted $\Psi_{j;k}$, where $j$ is the basis and $k$ the index of the
vector in that basis. The generalizations of the qubit Pauli
operators, denoted $X_d$ and $Y_d$, are defined by the actions
\begin{align}
X_d:& \Psi_{j;k} \rightarrow \Psi_{j;k+1}\nonumber\\ Y_d:& \Psi_{j;k}
\rightarrow \Psi_{j+1;k},
\end{align}
where additions are in modulo $d$ arithmetic. Each of the $N$
participants, denoted $n \in \{0,1,2,\cdots,N\}$, with the 0th player
being the secret dealer, applies the operation $X_d^{x_n} Y_d^{y_n}$
to the sequentially transmitted qudit, where $x_n, y_n \in
\{0,1,\cdots,d-1\}$ are private data held by the $n$th player. The
dealer measures in a random basis $\Xi$ to obtain outcome $\xi$.
Through public announcement, the players verify whether $\sum_{j=0}^n
y_n = \Xi$ (which happens with probability $1/d$) and reject the round
if not. If not rejected, then the protocol guarantees the perfect
correlations
\begin{equation}
\sum_{n=0}^N x_n = \xi,
\label{eq:req}
\end{equation}
which provides a basis for sharing a secret via a $(N,N)$~threshold
scheme \cite{THZ+15}. Over multiple rounds of the protocol, the
players can test for departure from (\ref{eq:req}), which can be used
to upper-bound eavesdropping. For details on security against
eavesdropping and or a collusional attack on the scheme, see
\cite{THZ+15}.

\subsection{Strong and weak quantum coin tossing\label{sec:CT}}

Suppose Alice and Bob are getting divorced. Coin tossing is a
crypto-task suitable to divide their assets randomly. Perfect coin tossing is a
primitive for Mental Poker \cite{GM82}. Classically,
(non-relativistic) coin tossing is based on computational complexity
\cite{Blu81}, whereas relativistic classical coin tossing involves Alice and Bob
sending each other messages with sufficient simultaneity as to ensure
the independence of their messages (see Section \ref{sec:RC}).

A crypto-task closely related to coin tossing is bit commitment, serves as
a primitive for other tasks such as zero-knowledge proofs
\cite{GMW91}. Bit commitment requires Alice to commit to a bit $a$ by submitting
an evidence to Bob. Later, she unveils $a$. During the holding phase
between commitment and unveiling, the evidence must bind Alice while
hiding $a$ from Bob. Quantum coin tossing can be built on top of quantum bit commitment: Alice commits to
a bit $a$; Bob publicly announces bit $b$; Alice unveils $a$. The toss
is $a + b ~\mod~2$. However, an unconditionally secure bit commitment protocol
cannot be constructed via calls to a secure coin tossing black box, even given
additional finite classical or quantum communication between Alice and
Bob, making bit commitment strictly stronger than coin tossing in the standard cryptographic
framework \cite{Kent1999coin}. 

It is conventionally believed that (nonrelativistic) quantum bit commitment
is not secure, owing to an entanglement-based attack uncovered by
Mayers, Lo and Chau (MLC) \cite{May97,LC97,CDP+13}, described briefly
below. For a similar reason, the impossibility of quantum coin tossing is also
accepted to hold generally \cite{DoK02}. Similar no-go arguments
exist for the impossibility of ideal coin flipping, oblivious transfer
and secure two-party computation.

The MLC argument can be cast as follows. In a quantum bit commitment protocol, suppose
$\rho_a$, with $a\in\{0,1\}$, denotes the density operator of the
evidence of commitment to bit $a$ that Alice submits to Bob. To be
concealing to Bob, we require
\begin{equation}
\rho_0 \simeq \rho_1. 
\label{eq:conceal}
\end{equation}
Mixed states can always be purified by going to a larger Hilbert
space. In this case, the purifications of $\rho_a$ must correspond to
Schmidt decompositions with the same Schmidt coefficients, say
$\xi_j$. We associate two of these purifications with the states
associated with Alice's commitment
\begin{align}
|\Psi_0\rangle &= \sum_j \sqrt{\xi_j}
|\alpha^A_j\rangle|\beta_j\rangle, \nonumber\\ |\Psi_1\rangle &=
\sum_j \sqrt{\xi_j} |\alpha^{A\prime}_j\rangle|\beta_j\rangle,
\label{eq:qbc}
\end{align}
where $|\beta_j\rangle$ are eigenstates of $\rho_0 = \rho_1$, while
the states $|\alpha^A_j\rangle$ and $|\alpha^{A\prime}_j\rangle$ are
orthogonal basis states. Alice can cheat because she only requires a
local rotation, connecting these two bases. She can use this local
unitary to switch her commitment between $|\Psi_0\rangle$ and
$|\Psi_1\rangle$ at the time of unveiling. This no-go result is an
application of the Hughston--Jozsa--Wootters theorem \cite{HJW93}, which
shows that any two ensembles with Bob having the same density
operator, can be prepared remotely by Alice, who holds the second
system that purifies Bob's state.

It may be noted that various authors have questioned the generality of
the cryptographic framework used to derive the standard no-go theorem
for quantum bit commitment \cite{RSri04, He11, He12, He14, Yue12, Che15}.

For the most part, the MLC result has motivated research in certain
directions deviating from ideal quantum bit commitment, among them bit string commitment
\cite{Ken03}, where Alice securely commits $n$ bits such that
recipient Bob can extract at most $m < n$ of these bits; a weaker form
of bit commitment, namely cheat-sensitive bit commitment, where the condition
(\ref{eq:conceal}) is relaxed to $\rho_0 \ne \rho_1$. Here, if either
party cheats, then the other party has a non-vanishing probability for
cheat detection \cite{HK04} (but see \cite{He15}). Note that even
cheat sensitive bit commitment is classically impossible.

Likewise, weaker versions of quantum coin tossing beyond ideal quantum coin tossing have been
studied. Here, one can distinguish between weak and strong flavors of
coin tossing. In strong coin tossing, the coin toss must be perfectly random. This is
the requirement of correctness. In weak coin tossing, it is known that the two
parties want opposite outcomes, e.g., Alice wants `heads' whereas Bob
`tails'. Then the coin tossing protocol need not protect against Alice biasing
the outcome towards `tails' and Bob towards `heads'. Strong coin tossing is
required when the bit preferred by the other party is not known or
their preferences clash.

The requirements for a (strong) quantum coin tossing protocol: 
\begin{description}
\item \emph{Correctness}. If both parties play honestly, then both outcomes
 of the coin are equal, namely $P(t=0)=P(t=1)=\frac{1}{2}$, where $t$
 is the toss outcome.
\item \emph{Bounded bias}. If Bob is honest but Alice is not, then $P_A(t)
 \le \frac{1}{2} + \epsilon_A$, where $\epsilon_A$ is her bias.
 Analogously for honest Alice and dishonest Bob, $P_B(t) \le
 \frac{1}{2} + \epsilon_B$.
\end{description}
The protocol's bias is $\epsilon$, defined as the maximum of
$\epsilon_j$ for $j \in \{A,B\}$. In an ideal quantum coin tossing, \mbox{$\epsilon_j = 0$}.
Quantum coin tossing can be based on quantum bit commitment \cite{SR01} or entanglement-based
\cite{Col07}.

Quantum coin tossing is known to offer an advantage over classical coin tossing in that quantum coin tossing can
guarantee a cheating probability strictly lower than 1, which is
impossible in a non-relativistic classical coin tossing scheme. Quantum coin tossing protocols
with bias $\epsilon$ lower than half have been proposed
\cite{ATV+00,Amb01,NS03,SR01}.

Kitaev \cite{Kit03} found that a bias lower than
$\frac{1}{\sqrt{2}}-\frac{1}{2} \approx 0.207$ cannot be achieved by a
fair (one where $\epsilon_A = \epsilon_B$) quantum coin tossing protocol, a bound that
has been demonstrated to be reachable arbitrarily close \cite{CK09}
(also cf. \cite{Moc05,Moc07}).

Quantum coin tossing under noisy conditions can draw help from quantum string coin
flipping \cite{LBA+05}. In \cite{HW11}, it is allowed for honest
players to abort with a certain probability determined by the level of
noise. Quantum coin tossing with partial noise tolerance by means of a nested
structure is proposed in \cite{ZZ15}. These considerations are
relevant to practical implementations of quantum coin tossing. Recent such works
include a string of coin tosses generated using photonic qutrits with
entanglement in orbital angular momentum \cite{MVU+05} and an all
optical-fiber single-coin quantum coin tossing \cite{NFH+08}. An experimental
realization of the loss resistant quantum coin tossing protocol proposed in
\cite{BBB+09} is reported in \cite{BBB+11}, where, however
entanglement is employed rather than a practical weak coherent source,
because of the protocol's vulnerability to multi-photon pulses. Quantum coin tossing,
which is primarily considered for two mistrustful parties, can be
generalized to multiple parties \cite{ABD+04}.

The coin tossing protocol \cite{BBB+09} uses the encoding states
\begin{equation}
|\chi_{\beta,a}\rangle = \sqrt{\frac{1+(-1)^a x}{2}}|0\rangle +
(-1)^{\beta+a}\sqrt{\frac{1-(-1)^a x}{2}}|1\rangle,
\label{eq:weird}
\end{equation}
where $x \in \{0,1\}$, $\beta$ represents the basis and $a$ the secret
bit. Alice partially commits to bit $a$ by submitting the states
$\rho_a = \frac{1}{2}(|\chi_{0,a}\rangle\langle\chi_{0,a}|+
|\chi_{1,a}\rangle\langle\chi_{1,a}|$. The supports of $\rho_0$ and
$\rho_1$ are not disjoint, and thus Bob's ability to discriminate
between $\rho_0$ and $\rho_1$ is constrained by a minimum error
discrimination bound. This is just the reason that Alice is able to
re-transmit a state if Bob's measurement fails and that the scheme has
loss-resistance in conjunction with the use of a single-photon source.

The protocol proposed in \cite{PCD+11} aims to correct this reliance
on a single-photon source (as against using a source of weak coherent
pulses), albeit by fixing the number of pulses emitted and thereby
bounding the multiphoton probability. However, its practical
realization \cite{PJL+14} is not found to be entirely loss-tolerant,
although admitting several imperfections that would be encountered in
practice.
 
It is an interesting question whether device-independent methods can be extended to
distrustful cryptography. It turns out that for quantum bit commitment with finite cheat
probability and bias, one can construct a device-independent scheme, and then build
coin tossing on top of that \cite{SCA+11}. These authors present a device-independent
scheme for cheat-sensitive quantum bit commitment, where Alice's and Bob's cheating
probabilities are $\simeq 0.854$ and 0.75, which is used to propose a
device-independent protocol for coin flipping with bias $\lesssim 0.336$.

Zhao \emph{et al} \cite{ZYW+15} report using measurement-device independence
\cite{LCQ12,XCQ+15} to protect quantum coin tossing against detector-side channel
attacks due to Alice, who may launch a detector-blinding attack based
on a recent experiment. This scheme essentially modifies the protocol
of \cite{BBB+09} to incorporate the measurement-device-independent method, but the authors also
consider the possibility of using weak coherent pulses. This scheme
is found to be loss-tolerant when single-photon sources are employed.
As expected from the use of measurement-device independence, the resulting measurement-device-independent quantum coin tossing is shown to
potentially offer a doubling of the secure distance in some cases.

\subsection{Quantum private query}

Private information retrieval or private query \cite{CKG+98} is a
crypto-task involving two parties, a user Alice interacting with a
server Bob, wherein Alice queries him to obtain an element held by Bob
in his private database, such that Bob does not know which element she
queried for (user security), while he in turn, wishes to restrict
information Alice may gain about other elements in the database
(database security). A protocol for quantum private query was proposed in
2007 \cite{GLM08,GLM10}, where it was shown to provide an exponential
savings in communicational and computational complexity \cite{AA15}.

While an unconditionally secure private query is known to be
impossible, practical, cheat-sensitive schemes can be proposed. The
basic idea of quantum private query can be illustrated using the phase-encoded scheme
proposed in \cite{O11}. Let server Bob possess $D$ elements in the
database, labelled $d(j) \in \{0,1\}$, where $0 \le j \le D-1$. To
query item $j$, Alice transmits the state $|\psi\rangle =
\frac{1}{\sqrt{2}}(|0\rangle + |j\rangle)$, whereas Bob performs the
oracle operation given by
\begin{equation}
U = \sum_{j=0}^{D-1} (-1)^{d(j)}|j\rangle\langle j|
\label{eq:oracle}
\end{equation}
whereby the query state transforms to
\begin{equation}
|\phi\rangle = \frac{1}{\sqrt{2}}(|0\rangle + (-1)^{d(j)}|j\rangle)
\end{equation}
and Alice determines her required information by distinguishing
between the two possibilities $\frac{1}{\sqrt{2}}(|0\rangle \pm
|j\rangle)$. Such a quantum private query protocol is of practical importance, assuming
Bob does not launch entanglement-based attacks.

In Eq.~(\ref{eq:oracle}), the oracle unitary $U$ can be difficult to
implement for large $D$. As a result, various quantum private query
protocols based on quantum key distribution have been proposed
\cite{JSG+11, GLW12, ZGG+13, YSX+14, WHZ11, CLM+14, YZY15, SLW+15,
 LGH+15}.

\subsection{Quantum fingerprinting and digital signatures}

Other related protocols include the quantum oblivious set-membership
\cite{SMZ+15} and private set intersection \cite{SMZ+16}. In
quantum oblivious set-membership, Bob's server decides if a user Alice's secret is a member of his
private set in an oblivious fashion, namely without his knowing which
element it is \cite{SMZ+15}. Requiring a communication cost of $O(1)$
bits, it yields an exponential reduction in communication cost with
respect to classical solutions to the problem.

Signature schemes, which are prevalent in today's electronic
communication, were first proposed by Diffie and Hellman in 1976 in
the classical framework. They permit messages to be transmitted from
a sender to multiple recipients, with the guarantee that the messages
remain authentic and can be forwarded to other recipients without
invalidation.

\pagebreak
In contrast to classical signature schemes, that depend on
computationally secure one-way protocols based on the
RSA algorithm or the elliptic curve digital
signature algorithm, a scheme for quantum digital signature
leverages quantum physical laws for the purpose.

In the first proposal for quantum digital signature \cite{GC01}, in analogy with the
classical signature scheme, a quantum public key is proposed,
which is a set of quantum states, while the private key is the
classical description of those states. A quantum one-way function thus
replaces the classical one-way function to guarantee unconditional or
information theoretic security. Note that quantum one-way or hash
functions have the further property that the quantum hashes can be
exponentially shorter than the original function input, thereby
yielding \emph{quantum fingerprints} \cite{BCW+01} (see \cite{Xu+15}
which reports an experimental realization).

In contrast to the preceding scheme for quantum digital signature, which required quantum
memory in order to hold the public key and were thus not practical, the authors of
\cite{DWA14,CDD14} propose a quantum digital signature scheme where this requirement is
absent, taking a giant stride towards practicality. A further
improvement on this is quantum digital signature protocols that have the same practical
requirements as quantum key distribution \cite{WDK+15}.

Quantum digital signature has been extended in analogy with its classical counterpart to
three or more parties \cite{AWA16}. From an experimental perspective,
both kilometer-range quantum digital signature \cite{DCK+16} as well as free-space quantum digital signature
\cite{CPH+16} have been demonstrated.

\subsection{Blind quantum computation}

Universal blind quantum computation is a measurement quantum
computation based protocol, wherein a quantum server carries out
quantum computation for client Alice, such that her input, output and
computation remain private and she does not require any memory or
computational power \cite{BFK09}. The protocol is interactive and has
a feed-forward mechanism whereby subsequent instructions by Alice to
the server can be based on single-qubit measurements. The method
submits naturally to fault tolerance.

Normally, the client must be able to prepare single-qubit states. But
even a classical client can perform blind quantum computation by interacting with two
entangled (but non-communicating) servers. It turns out that in this
setting, with authentication, any problem in bounded-error quantum polynomial time class has a two-prover
interactive proof with a classical verifier. Blind quantum computation has recently been
experimentally realized \cite{BKB+12}.

\section{Relativistic quantum cryptography \label{sec:RC}}

Unlike quantum key distribution, some mistrustful crypto-tasks are believed to be insecure
even when quantum resources are leveraged, among them, as we saw, bit
commitment and ideal coin tossing. Since bit commitment can act as a
primitive for various other crypto-tasks, such as zero-knowledge
proofs, these results are thought to weaken the case for the security
of quantum mistrustful protocols for communication and multiparty
computation.

However, these tasks may be secure under other frameworks, such as
that based on relativistic constraints or the assumption of noisy
storage with the adversary. Under the latter assumption, various
otherwise insecure two-party protocols become secure, among them
secure identification, oblivious transfer and quantum bit commitment
\cite{wehner2010implementation}.

A.~Kent \cite{Ken99} studied how bit commitment could be implemented by
exploiting special relativity constraints. Alice and Bob are
each split in two agents, and security is obtained against classical
attacks provided relativistic constraints can be invoked to prohibit
commucation between agents of the same player. The protocol evades
\cite{Hal04} the MLC attack \cite{May97,LC97} essentially by departing
from the concealment condition (\ref{eq:conceal}), but using
synchronous exchange of classical or quantum information between the
players in order to be concealing to Bob, which imposes strong
complexity, space and time constraints on the protocol.

This was followed by another scheme employing both quantum and
classical communication \cite{Ken12}, which was shown to be secure under
the assumption of perfect devices \cite{CK12,KTH+0}, and has been
experimentally realized as a robust method \cite{LKB+13,LCC+14}.
However, these protocols were restricted to a one-round communication,
which entails that for terrestrial agents, the commitment remains
valid for at most just over 20 ms. To improve on this, \cite{LKB+15}
proposed a method involving several rounds of classical communication,
which was proved secure against classical attacks, wherein the holding
phase could be made arbitrarily long via periodic, coordinated
communication between the agents of Alice and Bob. The bound on the
probability $\epsilon$ to cheat in this method was improved by other authors
independently \cite{chakraborty2015arbitrarily, fehr2016composition,
 pivoluska2016tight}. In particular, K.~Chakraborty \emph{et al}
\cite{chakraborty2015arbitrarily} show $\epsilon$ to satisfy the linear
bound: $\epsilon \lesssim (r+1)2^{(-n+3)/2}$, where $n$ is the length of the
bit string to be communicated between the agents at each of $r$
rounds. This allows the complexity of protocols to scale only
linearly with the commitment time, during which Alice and Bob are
required to perform efficient computation and communicate classically.

\pagebreak
Based on this theoretical breakthrough, E.~Verbanis \emph{et al} \cite{VMH+16} reported on a
relativistic bit commitment implementation for a 24-hour bit
commitment, with a potential for extension to over a year by modifying
the positions of agents. Recently, the possibility of making
relativistic quantum bit commitment device-independent has been studied \cite{AK15}. In
the case of quantum cryptographic tasks that are secure in the
relativistic setting, one can ask (as in bit commitment) whether special relativity by
itself can provide security, without invoking quantum mechanics (though quantum mechanics helps).

One crypto-task that requires a conjunction of both properties of
relativity and quantum mechanics is \emph{variable-bias coin toss}
\cite{CK06}, in which a random bit is shared by flipping a coin whose
bias, within a predetermined range, is covertly fixed by one of the
players, while the other player only learns the random outcome bit of
the toss. While one player is able to influence the outcome, the
other can save face by attributing a negative outcome to bad luck.
Security arises from the impossibility of superluminal signaling and
quantum theory.

Two other protocols, whose security is known to be guaranteed under
the conjunction of relativity and quantum mechanics are location-oblivious data
transfer \cite{kent2011location} and space-time-constrained
oblivious transfer \cite{Pit16}. The location-oblivious data
transfer involves two mistrustful parties, wherein Alice transfers data in Minkowski space to Bob at a
space-time location determined by their joint actions and that neither
can predict in advance. Alice is assured that Bob will learn the
data, while Bob is assured that Alice cannot find out the transfer
location. In the space-time-constrained
oblivious transfer, Bob has to output $a_b$ (see definition
of oblivious transfer above) within $B_b$, where $B_0$ and $B_1$ are spacelike
separated regions.

In contrast to bit commitment, some crypto-tasks, such as secure
two-party quantum computation of various classical functions
\cite{Col07,Lo97,BCS12}, in particular all-or-nothing oblivious
transfer \cite{Col07,Col09,Rud02} and 1-out-of-2 oblivious transfer \cite{Lo97},
which are believed to be insecure in nonrelativistic quantum settings,
remain so even in the context of relativistic quantum settings. In
1-out-of-2 oblivious transfer, Alice inputs two numbers $a_0$ and $a_1$, while Bob
inputs bit $b$ and then outputs $a_b$. In all-or-nothing oblivious transfer, Bob
retrieves a bit string sent by Alice with a probability half or gets
nothing at all. Also, position-based cryptography, which uses only
geographic position as the sole credential of a player, is known to be
insecure even with a conjunction of special relativity and quantum mechanics, if adversaries can
pre-share a quantum state of unbounded entanglement. A quantum
relativistic that is forbidden is that Alice can make available a
state received from Bob at an arbitrary event in the causal future, as
per the no-summoning theorem \cite{Ken13,HM16}.

\section{Technological issues\label{sec:tech}}

In this section, we cover the practical issues regarding experimental
realization of a quantum key distribution. This works in
tandem with advances in theory, for example, the quantum de Finetti
theorem, which would be applicable when it is difficult to bound the
dimension of the communication medium (possibly corrupted
maliciously). This result has been applied to derive secure quantum key distribution when
signals used are technologically limited to Gaussian states or weak
coherent states \cite{RC08}, rather than single-photons.

Practical challenges that emerge because of technological issues
include:
\begin{enumerate}
\item In discrete-variable protocols, key information is encoded in the polarization
 or the phase of weak coherent pulses simulating true single photon
 states. Hence, such implementations employ single photon detection
 techniques, e.g. BB84. However, the use of weak coherent pulses
 leads to some practical attacks such as the photon number splitting
 attack for which decoy states have to be used (cf. Section
 \ref{sec:pns}).
\item In the continuous variable protocols, information has to be
 encoded in the quadratures of the quantized electromagnetic fields
 such as those of the coherent states and homodyne or heterodyne
 detection techniques such as those used for optical classical
 communication (cf. Section \ref{sec:cvqkd}).
\item The security level of a protocol is decided by the type of
 attack considered in its security proof, which in turn could be
 dictated by technological considerations (e.g., Eve's ability to
 fight decoherence by realizing massive entangled states). Proving
 the security against collective (coherent) attacks and universal
 composability (which, for quantum key distribution, would cover joint attacks over the
 distribution of the key as well as its eventual use \cite{BHL+05}),
 at speeds and distance that are compatible with practical
 applications and technologically feasible, is quite a challenge. In
 practice, this would require the ability to realize efficient
 post-processing, including parameter estimation of quantum key distribution performance
 with stable setups across large data blocks. In a quantum network,
 the performance of any protocol is assessed point to point by
 considering the key distribution rate at a given security level
 under these attacks.
\end{enumerate}

For prevalent usage of quantum cryptography, low cost and robustness
are important. Among efforts being undertaken in this direction, it
has been shown that quantum key distribution systems can coexist with dense data traffic
within the same fibre, thereby precluding the need for dark fibers,
which are costly and moreover frequently unavailable
\cite{Pat14,DLB+16}. With access network architecture, multiple quantum key distribution
users can have simultaneous access in a way that would be compatible
with Gigabit Passive Optical Network traffic \cite{Fro15}. Yet
another direction to reduce not only cost, but also system complexity
and power consumption is through a chip-level photonic integration,
which would lead to a high degree of mass-manufacturable, low cost
miniaturization \cite{HNM+0}.

We first begin with a short introduction of classical fiber optical
communication \cite{GT98,Aga10} and then its adaptation for quantum
communication.

\subsection{Classical fiber-optics}

There has been a tremendous demand for increasing the capacity of
information transmitted and internet services. Scientists and
communication engineers are in pursuit of achieving this technological
challenge. The invention of LASER systems in the 1960s dramatically
altered the position of lightwave technologies as compared to radio or
microwaves. The availability of a coherent source allowed one to pack
an enormous amount of information into light signals increasing the
bandwidth. A lightwave communication system consists of a
transmission unit with source and electronics to modulate the signals,
an optical fiber channel connecting the sender and the receiver and
optical amplifiers (also known as repeaters) placed at certain
distances along the fiber link to boost the signal strength, a
receiving unit with optical detectors and accompanying electronics to
retrieve the original signal and extract the transmitted
information. Each unit of the fiber-optic communication system is
described briefly.

In standard telecom optical fibers of 1550 nm, attenuation of
light is 0.2 dB/km (improved in the recently developed ultralow-loss
fibers to 0.16 dB/km). This lossy property will restrict of
point-to-point quantum key distribution nodes to a few hundreds of kms and strong bounds on
the key rate \cite{TGW14,PLO+17}. With practical quantum key distribution, the rates
achieved are Mbit/s even though classical fiber optics can deliver
speeds upto 100 Gbit/s per wavelength channel.

\subsubsection{Transmission}

The choice of a source depends on the type of application. For
high-speed low loss communication with bit rates of the order of Gbps,
the source should meet the following requirements:\\
\\
\begin{enumerate}
\item Generation of wavelengths leading to low losses in the channel
 for a given power level such that the repeater spacing is large.
\item Spectral line width of the order of $\leq$ 1 nm to avoid
 dispersion (variation in phase velocity of a wave depending on its
 frequency).
\item High-speed modulation for achieving the desired transmission rate.
\end{enumerate}

Typically, semiconductor-based (InGaAsP or GaAs) light sources, such
as laser diodes and LEDs, are used in optical communication. They
emit required wavelengths, are highly efficient, compact in size and
can be modulated corresponding to the input electrical signals. 

LED diodes are basically forward-biased p-n junctions emitting
incoherent light due to spontaneous emission with 0.1~mW output
power and are suitable for transmission distances of a few kms at
10--100 Mbps bit rates. On the contrary, semiconductor laser diodes
emit coherent light via stimulated emission with an output power of
0.1~W suitable for longer distances at Gbps bit rates. Laser diodes
have narrow spectral-widths, allowing $50\%$ of the output power to be
coupled into fibers and useful in reducing chromatic dispersion. In
addition, laser diodes have a short recombination time, enabling them
to be directly modulated at high rates necessary for high-speed
long-distance communications.

High dimension quantum key distribution based on \mbox{$d$-level} systems allows transmission
of greater than 1 bit per photon detection, which can enhance
communication capacity at fixed particle rate
\cite{BT00,BKB01,CBK+02}. The round-robin differential phase shift quantum key distribution protocol (Section
\ref{sec:var}) allows a positive key in principle for any quantum bit
error rate \cite{SYK14}. Simply by choosing experimental
parameters, Eve's information can be tightly bounded, thereby removing
the need for monitoring the noise level of the channel. The strong
security of measurement-device-independent quantum key distribution is counterbalanced by the quadratic scaling of key
rate with detector efficiency, a drawback that can be overcome in
practice by reverting to detector-device-independent quantum key distribution (Section \ref{sec:ddi}).

\subsubsection{Channel} 

Optical fibers acting as transmission channels have a central
dielectric core (usually doped silica) with higher refractive index
surrounded by a cladding (pure silica) of lower refractive
index. Light signals are guided along the fiber axis using the
phenomenon of total internal reflection. Fibers with sudden and
gradual change in the refractive index at the core-clad boundary are
known as step-index (which include single-mode and multi-mode fibers)
and graded-index fibers respectively. Single-mode (multi-mode)
step-index fibers can sustain only one mode (many modes) of
light. Different modes travel at different speeds in a graded-index
fiber due to the gradual decrease in refractive index from the center
of the core, allowing all of them to reach the output at the same
instant of time, thereby reducing chromatic dispersion. 

Faithful transmission of signals through these channels depend on the
transmission characteristics of the fibers which include attenuation,
distortion, absorption, dispersion and scattering losses.
\begin{table*}
\begin{center}
\caption{\color[HTML]{000000} Comparison between single-mode and multi-mode step-index
 optical fibers. Note that the above mentioned transmission distance
 and rates are for classical communication.\\}
\begin{tabular}{| c |c|c|}
\hline
\textbf{Properties} & \textbf{Single-mode fibers} & \textbf{Multi-mode fibers} \\ 
\hline
Core ($\mu$m) & 8-10 & 50-100 \\
\hline
source & LASERs & LEDs \\
\hline
Transmission distance and rate & $> 1000$ km at 10Gbps & 550m at 10Gbps \\
\hline
Operating wavelengths (nm) & 1310 or 1500 & 850 or 1300 \\
\hline
Attenuation loss for above wavelenghts (dB/km) & 0.4 or 0.2 & 3 or 1 \\
\hline
Cost & high & low \\
\hline 
\end{tabular}

\label{table:fiber_prop}
\end{center}
\end{table*}

\subsubsection{Detection}

Optical detectors convert light signals into electrical signals which
is then amplified and processed by external circuity. Commonly used
detectors for fiber-optics are semiconductor-based using materials
such as Si, Ge, GaAs, InGaAs, owing to good response characteristics
in the optical domain and compatibility with optical fibers. Incident
light with energies greater than the bandgap of the semiconductor are
absorbed to generate e-h pairs leading to an external photocurrent.
The photocurrent is suitably amplified and processed for the
extraction of transmitted data. PIN (p-doped, intrinsic, n-doped
layers) diodes and Avalanche photodiodes (APDs) are mostly used for
photodetection. Both the devices are operated in the reverse-biased
condition and the e-h pairs are absorbed in the depletion region.

The key enabling factor of single-photon detectors is their low
noise, which in turn would depend on the type of the detection
technique. Room temperature single-photon detectors have been shown
to be suitable for high bit rate discrete-variable quantum key distribution \cite{C14}. For continuous
variable quantum key distribution (Section \ref{sec:cvqkd}), cooling is not necessary.

\subsection{Quantum communication\label{sec:co}}

With this background of classical communication, we now discuss
quantum communication using fiber-optics. Looking at the Table
\ref{table:fiber_prop} it is clear that single-mode fibers are
preferable for quantum communication. For secure quantum
communication, the sender and receiver are connected by quantum
channels. There is nothing special about these channels except for the
fact that the information is carried using single quantum systems
known as qubits, realized as photons, where information is encoded in
one of the degrees of freedom, in fact polarization.

Protecting the polarization of a photon from environmental effects
known as decoherence and decoupling the polarization degree of a
photon from its other degrees of freedom (such as frequency) to ensure
the faithful transmission of quantum information is very tricky.
Single-photons are fragile in nature and cannot sustain themselves
typically after traveling for 200 km.

Optical amplifiers known as quantum repeaters are placed at certain
intervals along the quantum communication network to maintain the
signal strength and increase the transmission distance. It is worth
noting here that, quantum repeaters \cite{SSM+07,SSR+11} are not a
straightforward extensions of their classical counterparts. Quantum
signals \textit{cannot} be detected or amplified directly without
disturbing it, by virtue of the \textit{no-cloning theorem}. Hence,
amplification and restoration of the original signal must be achieved
without direct interaction.

In addition, quantum cryptographic security requires the generation of
genuine random number sequences where each random number is completely
uncorrelated with other numbers in the sequence. It is
also not desirable to have any correlations across the runs among
different sequences. Quantum indeterminism forms the basis for
generation of truly random numbers. Measurement of a single quantum
system, an entangled state, coherent state, vacuum state are some
methods of random number generation. Quantum randomness cannot be
directly accessed at the macroscopic level. The quantum fluctuations
are classically amplified to extract genuine randomness (though there
is a theoretical proposal \cite{SSS0} for \textit{quantum}
amplification of quantum fluctuations). The random number sequences
generated are helpful in the random selection of basis for encoding a
qubit.

It is worth pointing out that measurement-device-independent quantum key distribution \cite{Cur14} is amenable
for upscaling to a multi-user, high speed communication networks in
metropolitan areas \cite{TYZ+16,C16}, inasmuch as measurement devices can
be positioned in an untrusted, dense relay, where is accessed by a
number of quantum key distribution users \cite{Fro0}, a scenario whose feasibility has been
validated by a number of groups (cf. \cite{Tan14} and references
therein), in particular discrete-variable measurement-device-independent quantum key distribution over a distance of 200 km in a
telecom fiber \cite{TYC+14} and 404 km in an ultralow-loss fiber
\cite{YCY+16}. Channel loss upto 60 dB can be tolerated given high
efficiency single-photon detectors, which translates to a distance of
300 km over standard telecom fiber \cite{Val15}. 

\emph{Quantum repeaters}. Photons are very fragile and hence for long-distance communication one
needs to maintain the signal to noise strength for faithful
communication \cite{BDC+98,DBC+99}. With quantum repeaters, the idea
is to divide the entire communication distance into smaller nodes with
quantum repeater stations, such that sufficiently noiseless
entanglement can be shared between two consecutive nodes. One then
performs entanglement swapping \cite{HBG+07} to entangle nodes farther
out, thereby establishing entanglement between far away nodes
\begin{align}
|\Psi\rangle_{1234} &= \frac{1}{2}(|00\rangle_{12}+|11\rangle_{12})
(|00\rangle_{34}+|11\rangle_{34}) \nonumber \\
&= (|\Phi^+\rangle_{14}|\Phi^+\rangle_{23} + |\Phi^-\rangle_{14}|\Phi^-\rangle_{23} 
\nonumber \\ &+
|\Psi^+\rangle_{14}|\Psi^+\rangle_{23} + |\Psi^-\rangle_{14}|\Psi^-\rangle_{23}),
\label{eq:entswap}
\end{align}
where $|\Psi^\pm\rangle = \frac{1}{\sqrt{2}}(|01\rangle \pm
|10\rangle)$ and $|\Phi^\pm\rangle = \frac{1}{\sqrt{2}}(|00\rangle \pm
|11\rangle)$ are the Bell states \cite{NC00}. In
Eq. (\ref{eq:entswap}) measuring particles 2-3 in a Bell basis
projects particles 1-4, that may never have interacted, into an
entangled state.

Based on the different
approaches to rectify fiber attenuation and operation (gate,
measurement) losses at each node and performance for specific
operational parameters (local gate speed, coupling efficiency, etc.),
one can classify the quantum repeaters into different generations
\cite{PL16, MKL+14, MJ}. Each generation aims to achieve better key
rates and decrease in memory errors for long-distance communication
\cite{LMK+15}. For loss (operational) error suppression, the method
employed is heralded generation (heralded purification) which is
probabilistic and involves two-way classical communication. But, the
quantum error correction approach for both is deterministic and
involves one-way communication. Various realizations of quantum
repeaters with or without memory are being explored
\cite{ATL15,SSR+11,S07,CTS+06,SDS09,ZDB12,MGH+14,RPL09}

\subsection{Single-photon sources}

Quantum communication, especially quantum cryptography and quantum
random number generation, demands that single-photons be employed
\cite{EFM+11,LO+05}, in order for standard security proofs such as
\cite{SP00,LC00} to work. Typically attenuated lasers are used as
substitute single-photon sources. Usually, they should emit photons
with mean photon-number $\mu=1$, variance $\Delta^2=0$ and their
second order correlation function $g^{(2)}(t)=0$. Ideally,
single-photon sources should generate single photons as and when
required, namely on-demand, with $100\%$ probability. Such
deterministic systems are of two types:

\begin{description}

\item \emph{Single emitters}. Single atoms, single ion and single molecule
 emitters are either $\Lambda$ or three-level systems in which
 controlling the pump laser and the atom-cavity coupling, a certain
 coherent state is transferred to the ground state via stimulated
 Raman adiabatic passage or radiative de-excitation respectively to
 generate a single photon in the cavity mode. These sources are
 scalable, emit indistinguishable photons, have low decoherence and
 multi-photon effects. Quantum dots \cite{BRV+12} and diamond
 nitrogen-vacancy (N-V) centers are other popular sources where single
 photons are generated by radiative recombination of electron-hole
 pairs and optical transitions in the N-V center respectively. But,
 they suffer from small coupling efficiency, scaling and
 indistingishability of the generated photons.

\item \emph{Ensemble-based emitters}. Single photons are generated by the
 collective excitations of atomic ensembles of Cs or Rb. The ensemble
 is also a $\Lambda$-type system with metastable ground states
 $|g_1\rangle$ and $|g_2\rangle$ and an excited state $|e\rangle$. A
 weak optical light is coupled to the population inverted atoms to
 induce the $|g_1\rangle \rightarrow |e\rangle$ transition. The
 de-excitation of a single photon from $|e\rangle \rightarrow
 |g_2\rangle$ is detected and its presence confirmed. This process is
 known as heralding. Next, a strong pulse induces a transition
 $|g_2\rangle \rightarrow |e\rangle$ generating a single heralded
 photon with $|e\rangle \rightarrow |g_1\rangle$ transition.

\end{description}

Single-photon sources based on probabilistic photon emission through
parametric down-conversion and four-wave mixing are also
available. The probability of multi-photon generation in such sources
increases with the probability of single-photon generation. A
single-photon detector cannot distinguish between single photons and
multiple photons. This imperfection can be used by an eavesdropper to
obtain secret key information after basis reconciliation by measuring
the photons acquired from these multi-photon pulses.

\subsection{Single-photon detectors}

An ideal single-photon detector should detect an incident photon with
$100\%$ probability and have nil dead-time and dark-count rates. There
are various types of single photon detectors (e.g., single-photon
avalanche photodiodes (InGaAs,Ge,Si), photo-multiplier tubes and
superconducting based nanowire single-photon detectors). However, none
of them can be considered as an ideal single photon detector as they
do not satisfy the above mentioned set of criteria that is expected to
be satisfied by an ideal single photon detector.

In particular, detection efficiency, wavelength dependence of
efficiency and dead time of single photon detectors are still a big
concern, and much effort has been made in the recent past to design
better detectors. Often the choice of optical components and the
frequency of transmission depend on the efficiency of the single
photon detector and the loss characteristics of the transmission
channel. Practically, it is an optimization.

The highest efficiency of single photon detectors is obtained for
incident photons of frequency around 800~nm, but the lowest
attenuation in an optical fiber happens around 1500~nm. Consequently,
open air quantum communication systems, including those which involve
satellites, are performed using photons with frequency near 800~nm, as
the single photon detectors perform best at this frequency, but
fiber-based implementations of quantum cryptography are realized in
teleportation range \mbox{(1350--1550~nm)}, where existing optical fibers show
minimum attenuation.

It is of cryptographic advantage if the detectors can also resolve the
number of photons in a pulse known as photon-number resolution.
Superconducting-tunnel-junctions, quantum dot optically gated field
effect transistors are some photon-number resolving detectors. Let us
discuss some of the detectors briefly. For a detailed comparison of
different detectors and their external circuitry refer \cite{KAA+00}.

\begin{description}
\item \emph{Photo-multiplier tubes (PMTs)}: An incident photon knocks an
 electron from a photocathode made of low work function material,
 which knocks more electrons causing an amplification of
 electrons. PMTs have large and sensitive collection areas, fast
 response time, about 10--40~$\%$ efficiency. They are vacuum
 operated which limits their scaling and integration abilities.
\item \emph{Single-photon avalanche photodiode (SPAD)}: An incident photon
 creates e-h pairs in the Geiger-mode operated photodiode
 \cite{CGL+04}. SPADs have a detection efficiency of $85\%$ but
 higher dark count rates as compared to PMTs. Also, once a pulse is
 detected, the wait time for re-biasing the circuitry for next
 detection, namely the dead time is longer. Schemes to reduce this
 afterpulsing have been realized. 
\item \emph{Quantum dot field effect transistors}: A thin layer of quantum
 dots between the gate and the conduction channel in a field-effect transistor traps
 incident photons modifying the channel conductance. This detector is
 useful for operation in the infrared region.
\end{description}

The above characteristics discussed are for non-photon resolving
operations but the detector's operation for photon-number resolution
is also being pursued. The active area of a detector is divided into
many pixels. Each pixel detects a photon and collectively many photons
are detected and resolved by the detector. Every time a pixel detects a
photon, the amplification process takes place independently and the
pixel undergoes dead- and recovery-time. Thus, the greater the number of
pixels, the better the resolution is.

\subsection{Photon-number splitting attacks \label{sec:pns}}

In quantum cryptography, the characteristics of
the single-photon sources and detectors dominate the practical
security issues. Multi-photon generation, blank pulses, detector
unresponsiveness for certain wavelengths, high dark counts, dead
times, recovery times and jitter are the crucial features which have
been used to launch powerful device attacks which cannot be detected
by usual methods.
In this context, we may specifically mention a particular type of
attack that arise due to our technological inability to build perfect
on-demand single photon source and photon number resolving detectors
that would not destroy the polarization states of the incident
photons. The attack is referred to as the photon number splitting
attack and illustrates a well known principle of cryptography---Alice and Bob are restricted by the available technology, but Eve is
not, she is restricted by laws of physics only (in other words, to
provide a security proof, we are not allowed to underestimate Eve by
assuming any technological limitations of the devices used by her).

Let us clarify the point. As we do not have a perfect on-demand single
photon source, we use approximate single photon sources, usually one
obtained by a weak laser pulse attenuated by a neutral density
filter. Such an approximate single photon source usually contains
single photon (in non-empty pulses), but with finite probability it
contains 2 photons, 3 photons, etc. Now, Eve may use a photon number
resolving detector to count the number of photons present in each
pulse (without changing the polarization state of the incident
photon), and stop all the single-photon pulses, while she allows all
the multi-photon pulses to reach Bob, keeping one photon from each
multi-photon pulse.

Subsequently, she may perform measurements on the photons that she
kept from the multi-photon pulses using right basis (based on Alice's
and Bob's announcements during basis reconciliation) without
introducing any disturbance. This is the photon number splitting attack, which requires a
photon-number resolving detector that does not destroy polarization
states of the incident photon. Although, quantum mechanics or any law
of physics does not prohibit construction of such a detector, until
now we do not have any technology to build such a detector.

Otherwise, Alice could use the similar strategy to the multi-photon
pulses and allow only single photon pulses to be transmitted. This
would have solve the need of single photon sources,
too. Unfortunately, no such photon number resolving detector exists
until now. However, we know a trick to circumvent photon number splitting attack, which
is the decoy state method \cite{Hwa03,LMC+05,Hwa05,BBW+16}.
Specifically, one may randomly mix intentionally prepared multi-photon
pulses (decoy qubits) with the pulses generated at the output of an
approximate single-photon source, which would generate single photon
pulses most of the time. Eve cannot selectively attack pulses
generated from the single photon source. In most incidents, pulses
originating from the single photon source will not reach Bob, but
those originating from the multi-photon source would reach Bob. Thus,
loss profile statistics for the pulses generated from the two sources
will be different and this difference (bias) would identify Eve, who
is performing photon number splitting attack from the natural channel noise which would
not be biased. Therefore, applying decoy states
\cite{Hwa03,LMC+05,Hwa05}, Alice and Bob can estimate both the
probability that a transmission results in a successful output as well
as the error rate for different initial pulses.

\subsection{Nonlinear effects}

Finally, we discuss some nonlinear effects that occur
in single-mode fibers that have an impact on its propagation
properties. Single-mode fibers are subject to polarization effects
such as birefringence (different phase velocities for orthogonal
polarization modes), polarization dependent losses (differential
attenuation between orthogonal modes) and polarization mode-dispersion
(different group velocities for orthogonal states). Fiber
irregularities and asymmetries are the cause for such effects which
can be overcome by polarization maintaining fibers where birefringence
is introduced on purpose to uncouple the polarization modes. Fibers
are also subject to dispersion, which is the broadening of signal
pulses in the time domain as they propagate along the fiber. Each
signal pulse consists different components which travel at different
speeds and hence their arrival time at the output varies.

In case of chromatic dispersion, different wavelengths travel at
different velocities. The overall chromatic dispersion in a fiber is
governed by the type of material used and its refractive index
profile. Since the material dispersion is fixed, the refractive index
profile has to be engineered in order to reduce such
effects. Dispersion compensating fibers and techniques (Bragg grating)
are employed to fix this issue.

\section{Continuous variable quantum cryptography \label{sec:cvqkd}}

Before we conclude this review, we need to mention that all the
single-photon-based schemes for quantum key distribution that are discussed here and most
of the other protocols for quantum key distribution, quantum secure direct communication and other cryptographic tasks
mentioned are discrete-variable based protocols in the sense that in these
schemes information is encoded in a discrete variable. However, it is possible to
implement most of these schemes by encoding information in continuous variable and
distributed phase reference, too \cite{SKJ+15}.

Basically, continuous-variable quantum key distribution involves homodyne detection instead of
photon counting encountered in discrete-variable quantum key distribution. Continuous-variable quantum key distribution was
first introduced with discrete modulation \cite{Hil00,Ral99,Rei00} and
later with Gaussian modulation using coherent states
\cite{GP02,GAW+03}.

Continuous-variable quantum key distribution and other continuous-variable based cryptographic schemes that are usually
implemented by continuous modulation of the light field quadratures
(usually with the coherent state \cite{GP02} or squeezed state
\cite{GP03, CLA01} of light), are important for various reasons. For
example, they are immune to some of the side-channel attacks that
exploit imperfections of single-photon detectors used in discrete-variable quantum key distribution to cause leakage of information. This is so because
coherent detectors (implementing homodyne or heterodyne detection) are
used in continuous-variable quantum key distribution.

Further, continuous-variable quantum key distribution can be implemented using commercially available
components \cite{QKA15} since the seminal work in continuous-variable quantum key distribution by Ralph in
1999 \cite{Ral99}. In this and the subsequent works by Ralph and his
collaborators \cite{Ral00}, small phase and amplitude modulations of
continuous wave light beams were exploited to carry the key
information. Subsequently, many schemes for continuous-variable quantum key distribution have been proposed
\cite{GP02, BP12, LKL04, GAW+03, QLP+15, SBL+15, PML+08, HHL+16} and
security proofs for a large set of those schemes have been provided
\cite{LG09, GP02, Lev15, FFB+12}, and interestingly some of the
security proofs are composable in nature (cf. \cite{Lev15} and
references therein).
Continuous-variable quantum key distribution has been experimentally realized by various groups
\cite{LBG+07, JKL+13}. For example, in \cite{LBG+07, JKL+13}
experimental realizations of long distance continuous-variable quantum key distribution has been reported.
However, continuous-variable quantum key distribution is not immune to all possible side channel attacks, and
various strategies to perform side channels attacks have been
discussed in the recent past (cf. \cite{HWY+13, QKA15, SKJ+15} and
references therein).

Although continuous-variable quantum key distribution protocols are not more complicated than their discrete-variable quantum key distribution
counterparts, the security analysis in continuous-variable quantum key distribution can be relatively
involved, with different considerations of hardware imperfections and
noise models. See the recent review \cite{LPF+17} and references
therein, where a less restricted notion of unconditional security in
continuous-variable quantum key distribution is considered. An earlier good overview covering the
conceptual issues but without detailed calculations is \cite{DL15}.

A composable security against general coherent attacks for continuous-variable quantum key distribution that
encodes via two-mode squeezed vacuum states and measurement by
homodyne detection, based on the uncertainty relation formulated in
terms of smooth entropies \cite{TR11}, is given in \cite{FFB+12}.
Also, see \cite{Geh15} (Section \ref{sec:1sdi}).

Continuous-variable quantum key distribution has been adapted to one-sided device-independent framework \cite{Geh15,
 walk2016experimental}, which would be relevant when secure hubs
(such as banks) are linked to less secure satellite stations. Continuous-variable quantum key distribution
has also been implemented in the measurement-device-independent quantum key distribution framework \cite{LZX+14,
 MSJ+14, Pir15}. Here, Charlie measures the correlation between two
Gaussian-modulated coherent states sent by Alice and Bob. However, continuous-variable
measurement-device-independent quantum key distribution requires homedyne detectors of efficiency over 85\% to
generate a positive key rate \cite{XCQ+15}, which has indeed been
recently attained \cite{Pir15,Geh15}. However, scaling up to an
optical network can be challenging because of losses in the detector
coupling and network interconnects (but see \cite{Pir15a}).
Therefore, in the measurement-device-independent quantum key distribution, for long distance communication, discrete-variable based
quantum key distribution is preferable to continuous-variable based, though the promise of high key rates
makes continuous-variable measurement-device-independent quantum key distribution an interesting option to consider. Techniques
proposed recently may help realize a dependable phase reference for
the continuous-variable quantum key distribution systems \cite{QLP+15,SBL+15,HHL+15}.
In a variant of this theme, quantum key distribution can also be based on continuous
variables such as spatial or temporal degrees of freedom, which are
basically used for upscaling the dimension of the information carrier
in quantum key distribution. The spatial degree of freedom of a photon can be used as the
information carrier, but this faces the technological challenge of
high-speed modulators being available \cite{Etch0,Mir15}.

Continuous variable quantum key distribution can be used to encode in large alphabets, such
as the arrival time of energy-time entangled photons \cite{ZSW08},
which was proven secure against collective attacks \cite{ZME+14} and
also realized experimentally, where it was found to achieve a rate of
6.9 bits per coincidence detection across a distance of 20 km at a key
rate of 2.7 MBits/s \cite{Zho15}. While this advancement improves the
key rate of entanglement based schemes vis-a-vis prepare-and-measure
quantum key distribution methods, practical implementation would require to meet the
challenge of attaining high level of interference visibility.

\section{Post-quantum cryptography}\label{sec:post-quantum}

Thus far, we have mentioned several schemes of quantum cryptopgraphy,
and noted that one of the main reasons behind the enhanced interest in
these schemes underlies in the pioneering work of Shor \cite{Sho94},
which entailed that if a scalabale quantum computer could be built,
then many classical schemes for key exchange, encryption,
authentication, etc., would not remain secure, as the quantum
algorithms are capable of performing certain computationally difficult
tasks (which are used to provide security in classical system) much
faster than their classical counterparts. Specifically, in a
post-quantum world (namely, when a scalable quantum computer will be
realized) RSA, DSA, elliptic curve DSA (ECDSA), ECDH, etc., would not
remain secure \cite{Che16}.

Here, we draw the reader's attention toward the point that ``quantum
algorithm can only perform certain computationally difficult tasks
(which are used to provide security in classical system) much faster
than their classical counterparts''. This is so because until now there
exist only a few quantum algorithms that provide the required speedup
(cf. \cite{Sho00} for an interesting discussion on ``Why haven't
more quantum algorithms been found?''). This leads to a few
questions---What happens to those classical cryptographic schemes
which use such computationally difficult problems that do not have a
quantum algorithm with required speedup? Can they be quantum
resistant in the sense that they can resist an adversary with a
scalable quantum computer?

Efforts to answer these questions led to a new branch of cryptography,
known as post-quantum cryptography that deals with families of
classical crytographic schemes which are expected to remain secure in
a world where practical, scalable quantum computers are available
\cite{Ber09}. Such schemes are usually classified into six families
\cite{Che16} such as: 
\begin{description}
\item \emph{Lattice-based cryptography}. This includes all cryptosystems
 that are based on lattice problems \cite{Pei+16, DR09}. These
 schemes are interesting as some of them are provably secure under a
 worst-case hardness assumption. However, it seems difficult to
 provide precise estimates on the security of these schemes against
 some well known techniques of cryptanalysis \cite{Che16}.
\item \emph{Code-based cryptography}. This encryption system is primarily
 based on error correcting codes. In these type of schemes, there is
 trade-off between key sizes and structures introduced into the
 codes. Added structures reduces key size \cite{BCG09}, but often
 allows attacks \cite{BLP08}. A classic example of this type is
 McEliece's hidden-Goppa-code public key encryption system, which was
 introduced in 1978 \cite{Mce78} and has not been broken until now
 \cite{Che16}.
\item \emph{Multivariate polynomial cryptography}. This is based on the
 computational difficulty associated in solving a set of multivariate
 polynomials over finite fields. Although, several schemes of this
 type have been broken \cite{DFP+07, Pat95}, confidence of the
 community is high on some of the schemes like Patarin's scheme for
 public-key-signature system that uses Hidden Fields Equations (HFE)
 and Isomorphisms of Polynomials (IP) \cite{Pat96}.
\item \emph{Hash-based signatures}. This includes schemes for digital
 signatures constructed using hash functions \cite{Ber09b, Moz15,
 Fan14}. Although, several hash-based systems have been broken in
 the past, confidence on the recent hash-based schemes is very high.
\item \emph{Secret-key cryptography}. Examples of type \emph{Advanced Encryption
 Standard} (AES), which is a symmetric private key encryption
 algorithm, created by Joan Daemen and Vincent Rijmen. A design goal
 behind AES is efficiency in software and hardware and software.
\item \emph{Other schemes} not covered under the above mentioned families.
\end{description}

Shor's algorithm cannot be used to attack the cryptosystems that
belong to above families as the associated computational tasks are
different. However, Grover's algorithm may be used to attack some of
the schemes, but since Grover's algorithm provides only a quadratic
speedup, an attack based on Grover's algorithm may be circumvented
using longer keys. Thus, it is believed that the schemes belonging to
above families would remain secure in the post-quantum world.

We have briefly mentioned about post-quantum cryptography, an
interesting facet of the modern cryptography as without a mention of
post-quantum cryptography any discussion on quantum cryptography would
remain incomplete. However, it is not our purpose to discus these
schemes in detail. We conclude this short discussion on post-quantum
cryptography by noting that the confidence of the cryptographic
community in these schemes is a bit artificial as it is impossible to
prove that faster quantum algorithms for all or some of the
computationally difficult problems used in these schemes will not be
designed in future. In brief, if a fast quantum algorithm for a task
is not available today, it does not mean that the same will not be
proposed tomorrow. Specifically, there is some practical reasons for
limited number of quantum algorithms that can provide required speedup
\cite{Sho00} and consequently, it is difficult to strongly establish
the security of the above mentioned schemes in the post-quantum world.

\section{Conclusions and perspectives\label{sec:conclu}}

In this brief review, we covered a number of quantum cryptographic
topics besides quantum key distribution, among them different crypto-tasks and
cryptographic frameworks. In a review of a vast area such as
quantum cryptography, it is, unfortunately, inevitable that some
important topics may not be covered. A case in point here is the
topic of quantum memory as applied to channel or device attacks. 

 Theoretically, the main work ahead in the area is in extending
security proofs in various scenarios to the composable framework under
the most general coherent attack. The main practical challenges are
perhaps developing on-chip quantum cryptographic modules that are free
from side channels and able to be scale to global networks, by
integrating point-to-point quantum cryptographic links. This may
drive the search for proper trade-offs between ease of implementation
and resource usage, or between reasonable security and economic
feasibility.

Regarding the foundational implications of quantum cryptography,
an interesting question is whether the no-go theorems that give
security to quantum cryptography can be used to derive quantum
mechanics. R.~Clifton \emph{et al} \cite{CBH03} presented a derivation of quantum mechanics from three
\emph{quantum cryptographic axioms}, namely, no-signaling, no-cloning and
no bit commitment. J.~Smolin \cite{Smo03} criticized this view by
presenting a toy theory that simulated these features but was not
quantum mechanics. In response, H.~Halvorson and J.~Bub \cite{HB03} argued that Smolin's toy
theory violated an independence reasonable condition for spacelike
separated systems assumed in \cite{CBH03}. More recently,
\cite{ASA16} have argued that general probability theories for single
systems can be distinguished between \emph{base} theories, which feature
a no-cloning theorem, which similar to Spekkens' toy theory that
defends an epistemic view of quantum states \cite{Spe07}, and
\emph{contextual} theories. The former supports a type of unconditional
security in the framework of trusted devices, whereas the latter
allows a degree of device independence. 

It is known that the usual definition of security in quantum key distribution implies
security under universal composition. However, keys produced by
repeated runs of quantum key distribution have been shown to degrade gradually. It would be
interesting to study direct secure communication (Section
\ref{sec:qsdc}) in the context of universal composability, and the
advantage of schemes for direct secure communication, if any, over quantum key distribution
under repeated usage.

\section*{Acknowledgments} 
Akshata Shenoy-Hejamadi acknowledges the support from Federal Commission for Scholarships
for Foreign Students through the Swiss Government Excellence
Postdoctoral Fellowship 2016--2017. Anirban Pathak thanks Defense Research and
Development Organization (DRDO), India for the support provided
through the project number ERIP/ER/1403163/M/01/1603. He also thanks
Kishore Thapliyal and Chitra Shukla for their interest and feedback on
this work.

\end{document}